\pdfoutput=1
\documentclass[cernpreprint,texlive=2016,txfonts,UKenglish]{latex/atlasdoc}

\usepackage[biblatex=true]{latex/atlaspackage}
\usepackage{latex/atlasphysics}

\newcommand{\AtlasCoordFootnote}{
  ATLAS uses a right-handed coordinate system with its origin at the
  nominal interaction point (IP) in the centre of the detector and the
  $z$-axis along the beam pipe. The $x$-axis points from the IP to the
  centre of the LHC ring, and the $y$-axis points upwards. Cylindrical
  coordinates $(r,\phi)$ are used in the transverse plane, $\phi$
  being the azimuthal angle around the $z$-axis. The pseudorapidity is
  defined in terms of the polar angle $\theta$ as $\eta = -\ln
  \tan(\theta/2)$. 
}

\graphicspath{{logos/}}

\usepackage{latex/atlasbiblatex}

\def\ppg{pp\rightarrow\gamma\ +\ X}
\def\pt{p_{\mathrm{T}}}
\def\R138{R^{\gamma}_{13/8}}
\def\Z138{R^{Z}_{13/8}}
\def\D138{D^{\gamma/Z}_{13/8}}
\def\etg{E_{\mathrm{T}}^{\gamma}}
\def\etag{\eta^{\gamma}}
\def\etiso{E_{\mathrm{T}}^{\mathrm{iso}}}
\def\etaphi{\eta\text{--}\phi}
\def\etisocut{$4.2\cdot 10^{-3}\cdot\etg+4.8\ {\mathrm{\GeV}}$}
\def\dsetg{{\mathrm{d}}\sigma/{\mathrm{d}}\etg}
\def\dsZ{\sigma^{\mathrm{fid}}_{Z}(13\ {\mathrm{\TeV}})/\sigma^{\mathrm{fid}}_{Z}(8\ {\mathrm{\TeV}})}
\def\as{\alpha_{\mathrm{s}}}
\def\asz{\alpha_{\mathrm{s}}(\mz)}
\def\mz{m_{\mathrm{Z}}}
\def\sher{{\textsc{Sherpa}}}
\def\pyt{{\textsc{Pythia}}}
\def\jetp{{\textsc{Jetphox}}}
\def\dytub{{\textsc{Dyturbo}}}
\def\dynn{{\textsc{Dynnlo}}}
\def\muR{\mu_{\mathrm{R}}}
\def\muF{\mu_{\mathrm{F}}}
\def\muf{\mu_{\mathrm{f}}}
\def\pb1{pb$^{-1}$}
\def\fb1{fb$^{-1}$}
\def\ele{e^+e^-}

\mathchardef\mhyphen="2D

\AtlasTitle{Measurement of the ratio of cross sections for inclusive
isolated-photon production in $pp$ collisions at $\sqrt s=13$ and
$8$~\TeV\ with the ATLAS detector}

\author{The ATLAS Collaboration}

\AtlasJournalRef{JHEP 04 (2019) 093}
\AtlasDOI{10.1007/JHEP04(2019)093}
\PreprintIdNumber{CERN-EP-2018-340} 

\AtlasAbstract{
The ratio of the cross sections for inclusive isolated-photon production in 
$pp$ collisions at centre-of-mass energies of 13 and 8~\TeV\ is measured 
using the ATLAS detector at the LHC. The integrated luminosities of the 
13~\TeV\ and 8~\TeV\ datasets are 3.2~\fb1\ and 20.2~\fb1, respectively.
The ratio is measured as a function of the photon transverse energy in
different regions of the photon pseudorapidity. The predictions from
next-to-leading-order perturbative QCD calculations are compared with
the measured ratio. The experimental systematic uncertainties as well
as the uncertainties affecting the predictions are evaluated taking
into account the correlations between the two centre-of-mass energies,
resulting in a reduction of up to a factor of $2.5$ ($5$) in the
experimental (theoretical) systematic uncertainties. The predictions
based on several parameterisations of the proton parton distribution
functions agree with the data within the reduced experimental and
theoretical uncertainties. In addition, this ratio to that of the
fiducial cross sections for $Z$ boson production at 13 and 8~\TeV\
using the decay channels $Z \rightarrow e^+e^-$ and $Z \rightarrow
\mu^+\mu^-$ is made and compared with the theoretical predictions. In
this double ratio, a further reduction of the experimental uncertainty
is obtained because the uncertainties arising from the luminosity
measurement cancel out. The predictions describe the measurements of
the double ratio within the theoretical and experimental uncertainties.
}

\def\figdir{figures/}

\begin{document}

\maketitle

\tableofcontents

\section{Introduction}
\label{intro}

The production of prompt photons in proton--proton collisions, $\ppg$,
provides a means of testing perturbative QCD (pQCD) with a hard
colourless probe. Since the dominant production mechanism in $pp$
collisions at the LHC proceeds via the $qg\rightarrow q\gamma$
process, measurements of prompt-photon\footnote{All photons produced
  in $pp$ collisions that are not secondaries from hadron decays are
  considered to be ``prompt''.} production are sensitive to the gluon
density in the proton~\cite{np:b860:311,epl:101:61002}. These
measurements can also be used to tune Monte Carlo (MC) models to
improve our understanding of prompt-photon production and aid those
analyses for which events containing photons are an important
background.

At leading order (LO) in pQCD, two processes contribute to
prompt-photon production: the direct-photon process, in which the
photon originates directly from the hard interaction, and the
fragmentation-photon process, in which the photon is emitted in the
fragmentation of a high transverse momentum ($\pt$)
parton~\cite{pr:d76:034003,pr:d79:114024}.

Measurements of prompt-photon production at a hadron collider
necessitate an isolation requirement to reduce the large contribution
of photons from hadron decays and the fragmentation component
in which the emitted photon is close to a jet.  The production of
isolated photons in $pp$ collisions has been measured previously by
the
ATLAS~\cite{pr:d83:052005,pl:b706:150,pr:d89:052004,jhep:1608:005,pl:b770:473}
and CMS~\cite{prl:106:082001,pr:d84:052011} collaborations at
centre-of-mass energies (\!$\sqrt s$) of $7$, $8$ and $13$~\TeV.

Comparisons of measurements of prompt-photon production and pQCD
predictions are usually limited by the theoretical uncertainties
associated with the missing higher-order terms in the perturbative
expansion. The measurements of inclusive isolated-photon cross
sections performed by ATLAS at $13$~\TeV~\cite{pl:b770:473} and
$8$~\TeV~\cite{jhep:1608:005} were compared with the predictions of
pQCD at next-to-leading order
(NLO)~\cite{jhep:0205:028,pr:d73:094007}. At both centre-of-mass
energies, the uncertainties affecting the predictions are dominated by
terms beyond NLO and are larger than those of experimental nature,
preventing a more precise test of the theory. An avenue to reach a
more stringent test is the inclusion of next-to-next-to-leading-order
(NNLO) QCD corrections in the
calculations~\cite{prl:118:222001}. Another avenue is to make
measurements of the ratio of cross sections for inclusive
isolated-photon production at $13$ and $8$~\TeV\ ($\R138$) and compare
them with the predictions~\cite{jhep:1208:010,epj:c78:470}. The impact
of the experimental systematic uncertainties and theoretical
uncertainties on the ratio of the cross sections is reduced, allowing
a more precise comparison between data and theory. This is achieved by
accounting for inter-$\!\sqrt s$ correlations in the experimental
systematic uncertainties affecting the measurements and in the
uncertainties of the theory predictions.

A further reduction of the experimental uncertainty can be achieved by
measuring a double ratio: the ratio of $\R138$ to the ratio of the
fiducial cross sections for $Z$ boson production at $13$~\TeV\ and
$8$~\TeV\ ($\Z138\equiv\dsZ$) presented in Ref. \cite{jhep:1702:117}.
The measurements of the fiducial cross sections  for $Z$ boson
production use the decay channels $Z \rightarrow e^+e^-$ and $Z
\rightarrow \mu^+\mu^-$. This observable, $\D138\equiv\R138/\Z138$,
can be viewed as the increase of the cross section for isolated-photon
production as a function of $\sqrt s$ normalised to the increase for
$Z$ boson production as a function of $\sqrt s$. Measuring $\D138$ is
beneficial because the uncertainties from the luminosity measurement
cancel out, and $\D138$ has only a slightly larger theory uncertainty
than $\R138$.

This paper presents measurements of the ratio of cross sections for
isolated-photon production in $pp$ collisions at $\sqrt s=13$~\TeV\
and $8$~\TeV\ with the ATLAS detector at the LHC. The phase-space
region is given by the overlap of the ATLAS measurements at $\sqrt
s=13$ and $8$~\TeV, defined by the photon transverse
energy\footnote{\AtlasCoordFootnote}  ($\etg$) in the range
$\etg>125$~\GeV\ and the photon pseudorapidity ($\etag$) in the region
$|\etag|<2.37$, excluding the region $1.37<|\etag|<1.56$. The photon
is isolated by requiring that the transverse energy inside a cone of
size $\Delta R \equiv \sqrt{(\Delta\eta)^{2} + (\Delta\phi)^{2}}=0.4$
in the $\etaphi$ plane around the photon direction, $\etiso$, is
smaller than
$E_{\mathrm{T,cut}}^{\mathrm{iso}}\equiv$~\etisocut~\cite{jhep:1608:005,pl:b770:473}.
Non-isolated prompt photons are not considered as signal. The
measurements of the ratios are based on the ATLAS measurements at
$13$~\TeV~\cite{pl:b770:473} and $8$~\TeV~\cite{jhep:1608:005} and a
detailed study of the correlations of the experimental systematic
uncertainties between the two centre-of-mass energies is presented
here. The measurement of the ratios is presented as a function of
$\etg$ in different regions of $\etag$, namely $|\etag|<0.6$,
$0.6<|\etag|<1.37$, $1.56<|\etag|<1.81$ and
$1.81<|\etag|<2.37$. Next-to-leading-order pQCD predictions for the
ratio are compared with the measurements. In addition, measurements of
$\D138$ are presented using the ATLAS results for
$\Z138$~\cite{jhep:1702:117}; the measurements are compared with
available theory predictions.

The paper is organised as follows: the ATLAS detector is described in
Section~\ref{detector}. The analysis strategy is summarised in
Section~\ref{analysis}. Fixed-order QCD predictions and their
uncertainties are discussed in Section~\ref{nlo}. Section~\ref{syst}
is devoted to the description of the experimental uncertainties. The
results are reported in Section~\ref{results}. A summary is given in
Section~\ref{summary}.

\section{ATLAS detector}
\label{detector}

The ATLAS experiment~\cite{PERF-2007-01} at the LHC uses a
multipurpose particle detector with a forward--backward symmetric
cylindrical geometry and a near $4\pi$ coverage in solid angle. It
consists of an inner tracking detector, electromagnetic (EM) and
hadronic calorimeters, and a muon spectrometer. The inner detector is
surrounded by a thin superconducting solenoid and includes silicon
detectors, which provide precision tracking in the pseudorapidity
range $|\eta|<2.5$, and a transition-radiation tracker providing
additional tracking and electron identification information for
$|\eta|<2.0$. For the $\sqrt s = 13$~\TeV\ data-taking period, the
inner detector also includes a silicon-pixel insertable
B-layer~\cite{jinst:13:T05008,ATLAS-TDR-19}, providing an additional
layer of tracking information close to the interaction point. The
calorimeter system covers the range $|\eta|<4.9$. Within the region
$|\eta|<3.2$, EM calorimetry is provided by barrel and endcap
high-granularity lead/liquid-argon (LAr) EM calorimeters, with an
additional thin LAr presampler covering $|\eta|<1.8$ to correct for
energy loss in material upstream of the calorimeters; for $|\eta|<2.5$
the LAr calorimeters are divided into three layers in depth. Hadronic
calorimetry is provided by a steel/scintillator-tile calorimeter for
$|\eta|<1.7$ and two copper/LAr hadronic endcap calorimeters for
$1.5<|\eta|<3.2$. The forward region is covered by additional
coarser-granularity LAr calorimeters up to $|\eta|=4.9$. The muon
spectrometer consists of three large superconducting toroidal magnets,
one barrel and two endcaps, each containing eight coils, precision
tracking chambers covering the region $|\eta|<2.7$, and separate
trigger chambers up to $|\eta|=2.4$. For the data taken at $8$~\TeV, a
three-level trigger system was used. The first-level trigger was
implemented in hardware and used a subset of the detector
information. This was followed by two software-based trigger levels
that together reduce the accepted event rate to approximately
$400$~Hz. For the data taken at $13$~\TeV, the trigger was
changed~\cite{epj:c77:317} to a two-level system, using custom
hardware followed by a software-based level which runs offline
reconstruction software, reducing the event rate to approximately
$1$~kHz.

\section{Analysis strategy}
\label{analysis}

The measurements of ratios of cross sections presented in this paper
are based on the measurements presented in previous ATLAS
publications~\cite{jhep:1608:005,pl:b770:473,jhep:1702:117}, where
details of the analyses are given. The strategies followed for the
measurement of the ratios and for the theoretical predictions are
described below.

\subsection{Analysis strategy for $\R138$}
The measurements of $\dsetg$ at $\sqrt s=8$~\TeV\ ($13$~\TeV) used in
the measurement of $\R138$ are based on an integrated luminosity of
$20.2\pm 0.4$~\fb1\ ($3.16\pm 0.07$~\fb1). The measurement of the
ratio covers the range $\etg>125$~\GeV\ and is performed separately in
the four regions of $\etag$ defined in Section~\ref{intro}. A summary
of the analyses leading to the measurements of the differential cross
sections for inclusive isolated-photon production at $\sqrt s=13$ and
$8$~\TeV\ is given below.

Photon candidates are reconstructed from clusters of energy deposited
in the EM calorimeter. Candidates without a matching track or
reconstructed conversion vertex in the inner detector are classified
as unconverted photons, while those with a matching reconstructed
conversion vertex or a matching track consistent with originating from
a photon conversion are classified as converted
photons~\cite{epj:c76:666}. The photon identification is based
primarily on shower shapes in the calorimeter~\cite{epj:c76:666}. It
uses information from the hadronic calorimeter, the lateral shower
shape in the second layer of the EM calorimeter and the shower shapes
in the finely segmented first EM calorimeter layer to ensure the
compatibility of the measured shower profile with that originating
from a single photon impacting the calorimeter. The photon energy
measurement is made using calorimeter and, when available, tracking
information. An energy calibration~\cite{epj:c74:3071} is applied to
the candidates to account for upstream energy loss and both lateral
and longitudinal leakage. Events with at least one photon candidate
with calibrated $\etg > 125$~\GeV\ and $|\etag|<2.37$ excluding the
region $1.37<|\etag|<1.56$ are selected. The isolation transverse
energy $\etiso$ is corrected for leakage of the photon energy into the
isolation cone and the estimated contributions from the underlying
event (UE) and additional inelastic $pp$ interactions (pile-up). The
latter two corrections are computed simultaneously on an
event-by-event basis using the jet-area
method~\cite{jhep:0804:005,jhep:1004:065}. After these corrections,
isolated photons are selected by requiring $\etiso$ to be lower than
$E_{\mathrm{T,cut}}^{\mathrm{iso}}$. A small background contribution
still remains after imposing the photon identification and isolation
requirements and is subtracted using a data-driven method based on
background control regions~\cite{pl:b770:473,jhep:1608:005}. The
selected samples of events are used to unfold the distribution in
$\etg$ for each $|\etag|$ region to a phase-space region close to that
used for event selection.

The phase-space region at particle level uses particles with a decay
length $c\tau >$~10~mm; these particles are referred to as
``stable''. The particle-level isolation requirement for the photon is
built by summing the transverse energy of all stable particles, except
for muons and neutrinos, in a cone of size $\Delta R=0.4$ around the
photon direction after the contribution from the UE is subtracted; the
same subtraction procedure and isolation requirement used on data are
applied at the particle level.

An important part of this analysis is the evaluation of the
experimental systematic uncertainties in the ratio of the cross
sections at $13$ and $8$~\TeV\ taking into account correlations. This
study is described in Section~\ref{syst}. Given the dominance of the
systematic uncertainty arising from the photon energy scale when
measuring the cross sections, it is necessary to carefully study this
source of uncertainty. This source of systematic uncertainty is
decomposed into independent components~\cite{epj:c74:3071} and the
treatment of the correlations of these components between the
measurements at $13$ and $8$~\TeV\ results in a reduction of the
systematic uncertainty of the ratio.

The measurements of the ratio of cross sections are compared with NLO
pQCD predictions for which a proper evaluation of the theoretical
uncertainties is also of importance. The theoretical uncertainties in
the predictions for the cross sections are ${\cal O}$(10--15\%) for
both centre-of-mass energies and are dominated by contributions from
terms beyond NLO. These uncertainties are much larger than those of
experimental nature and limit how precisely the predictions can be
tested. The study of the theoretical uncertainties in the ratio is
described in Section~\ref{nlo}. As is the case for the experimental
systematic uncertainties, it is imperative that for each source of
theoretical uncertainty the degree of correlation between the two
centre-of-mass  energies is taken into account. As a result, the
theoretical uncertainty is reduced in the ratio, thus allowing a more
stringent test of the predictions.

\subsection{Analysis strategy for $\D138$}
\label{analyd138}
The measurement of the double ratio $\D138$ is based on the
measurement of $\R138$ described above as well as on the measurement
of $\Z138$. It should be noted that $\R138$ is measured as a function
of $\etg$ in different ranges of $\etag$, while $\Z138$ is a single
number. The measurement of $\Z138$ used here is the one reported in
Ref.~\cite{jhep:1702:117}. The fiducial cross section at a given
$\sqrt s$, $\sigma_{Z}^{\mathrm{fid}}(\sqrt s)$, is defined as the
production cross section of a $Z$ boson times the branching ratio of
the decay into a lepton pair of flavour $\ell^+\ell^- = e^+e^-$ or
$\mu^+\mu^-$ within the following phase space: the lepton transverse
momentum $p_\mathrm{T}^{\ell}> 25$~\GeV, the lepton pseudorapidity
$|\eta_{\ell}|<2.5$ and the dilepton invariant mass $66 < m_{\ell\ell}
< 116$~\GeV. The measurement at $\sqrt s = 13$~\TeV\ was performed in
the aforementioned phase space while the measurement at $\sqrt s =
8$~\TeV\ was extrapolated to the same phase space as described in
Ref.~\cite{jhep:1702:117}. Measurements of the fiducial cross sections
were made using the decay channels $Z \rightarrow e^+e^-$ and $Z
\rightarrow \mu^+\mu^-$, and combined for the final result. The
measured $\Z138$ is $1.537\pm 0.001\ ({\mathrm{stat.}})\pm 0.010\
({\mathrm{syst.}})\pm 0.044\ ({\mathrm{lumi.}})$~\cite{jhep:1702:117},
where ``stat.'' denotes the statistical uncertainty, ``syst.'' denotes
the systematic uncertainty and ``lumi.'' denotes the uncertainty due
to the ratio of the integrated luminosities. The evaluation of the
systematic uncertainty in the ratio takes into account correlations of
systematic uncertainties across channels and $\sqrt s$ as described in
Ref. ~\cite{jhep:1702:117}.

The predictions for $\D138$ are obtained from NLO pQCD calculations
for $\R138$~\cite{jhep:0205:028,pr:d73:094007} and NNLO pQCD
calculations for $\Z138$~\cite{prl:98:222002,prl:103:082001}. The
evaluation of the uncertainties affecting the predictions for $\D138$
requires considerations that account for the correlations arising from
the parton distribution functions (PDFs) and the strong coupling
constant, $\asz$.

\section{Fixed-order QCD predictions}
\label{nlo}

The theoretical predictions for the ratios of cross sections are
obtained using fixed-order QCD calculations. Details of the generators
and of the estimations of the theoretical uncertainties are given
below, especially emphasising the correlations between the two
centre-of-mass energies.

\subsection{Theoretical predictions for $\R138$}
\label{nlor}
The theoretical predictions for $\R138$ presented here are based on
NLO QCD calculations computed using the program
\jetp~1.3.1\_2~\cite{jhep:0205:028,pr:d73:094007}. This program
includes a full NLO QCD treatment of both the direct- and
fragmentation-photon contributions to the cross section for the $\ppg$
reaction. The number of quark flavours is set to five. The
renormalisation ($\muR$), factorisation ($\muF$) and fragmentation 
($\muf$) scales are chosen to be $\muR=\muF=\muf=\etg$. The
calculations are performed using various parameterisations of the
proton PDFs and the BFG set II of parton-to-photon fragmentation
functions at NLO~\cite{epj:c2:529}. The nominal calculation is based
on the MMHT2014 PDF set~\cite{epj:c75:204}. Predictions are also
obtained with other PDFs, namely  CT14~\cite{pr:d93:033006},
HERAPDF2.0~\cite{epj:c75:580}, NNPDF3.0~\cite{jhep:1504:040} and
ABMP16~\cite{pr:d96:014011}. For MMHT2014, CT14, HERAPDF2.0 and
NNPDF3.0 parameterisations of the PDFs, the sets determined at NLO are
used. For ABMP16, the set at NNLO is used. The strong coupling
constant $\asz$ is set to the value assumed in the fit to determine
the PDFs; as an example, in the case of MMHT2014 PDFs, $\asz$ is set
to the value $0.120$.

The calculations are performed using a parton-level isolation criterion
for the photon, which requires a total transverse energy of the partons
inside a cone of radius $R=0.4$ around the photon direction below
$E_{\mathrm{T,cut}}^{\mathrm{iso}}$. The predictions from \jetp\
are at parton level,\footnote{The parton level in \jetp\ consists of
  the generated photon and the few partons simulated with the matrix
  elements, while in \pyt\ it includes the partons after the parton
  shower.} while the measurements are at particle level. Corrections
for the non-perturbative (NP) effects of hadronisation and the UE are
estimated using samples from \pyt~8.186~\cite{cpc:178:852} as
described below. First, a correction factor ($C^{\mathrm{NP}}_{\sqrt
  s}$) is derived for the isolated-photon cross section at each
centre-of-mass energy as the ratio of the cross section at particle
level for a \pyt\ sample with UE effects to the \pyt\ cross section at
parton level without UE effects. Second, the ratio of the correction
factor for $\sqrt s=13$~\TeV\ to that for $\sqrt s=8$~\TeV,
$C^{\mathrm{NP}}_{\mathrm{R}}=C^{\mathrm{NP}}_{13}/C^{\mathrm{NP}}_{8}$,
is evaluated. The ratio of correction factors is obtained using the
ATLAS set of tuned parameters A14~\cite{ATL-PHYS-PUB-2014-021} with
the LO NNPDF2.3 PDF set~\cite{np:b867:244}. The ratio of correction
factors for non-perturbative effects applied to the ratio predictions
from \jetp\ is $C^{\mathrm{NP}}_{\mathrm{R}}= 0.9964 \pm
0.0020$.

The following sources of uncertainty in the theoretical predictions
are considered:
\begin{itemize}
\item The uncertainty in the NLO QCD predictions due to terms beyond
  NLO is estimated by repeating the calculations using values of
  $\muR$, $\muF$ and $\muf$ scaled by the factors $0.5$ and $2$. The
  three scales are either varied simultaneously or individually; in
  addition, configurations in which one scale is fixed and the other
  two are varied simultaneously are also considered. In all cases, the
  condition $0.5\leq\mu_A/\mu_B\leq 2$ is imposed, where
  $A,B= \textrm{R, F, f}$. The final uncertainty is taken as the
  largest deviation from the nominal value among the $14$ possible
  variations.
\item The uncertainty in the NLO QCD predictions related to the
  proton PDFs is estimated by repeating the calculations using the
  $50$ additional sets from the MMHT2014 error analysis.
\item The uncertainty in the NLO QCD predictions related to the
  value of $\asz$ is estimated by repeating the calculations using two
  additional sets of proton PDFs from the MMHT2014 analysis for which
  different values of $\asz$ were assumed in the fits, namely
  $\asz=0.118$ and $0.122$~\cite{epj:c75:435}.
\item The impact of the beam energy uncertainty is
  estimated by repeating the calculations with $\sqrt s$ varied by its
  uncertainty of $0.1\%$~\cite{prab:20:081003}.
\item The uncertainty in the corrections for non-perturbative effects
  is estimated by comparing the results of using variations of the A14
  tune in which the parameter settings related to the modelling of the
  UE are changed~\cite{ATL-PHYS-PUB-2014-021}. 
\end{itemize}

For the individual differential cross sections and for both
centre-of-mass energies, the dominant theoretical uncertainty arises
from the estimate of contributions from terms beyond
NLO~\cite{jhep:1608:005,pl:b770:473}.

The predictions for $\R138$ are obtained by calculating the ratio of
the individual differential cross sections at each centre-of-mass
energy. To estimate the theoretical uncertainty in $\R138$, the
correlation between the two centre-of-mass energies for each source
listed above needs to be considered. The uncertainties due to the
PDFs, $\asz$, beam energy and non-perturbative effects are fully
correlated between  the two centre-of-mass energies. The relative
uncertainties in $\R138$ due to the uncertainties in $\asz$, the PDFs
and the beam energy exhibit a significant degree of cancellation with
respect to the individual predictions. However, for the scale
uncertainties, the correlation is a priori unknown. In the standard
approach, varying the scales coherently or incoherently at both
centre-of-mass energies leads to very different theoretical
uncertainties:
\begin{itemize}
 \item In the coherent case, there are large cancellations in the
   uncertainties in the predictions for $\R138$, particularly in the
   variation of $\muR$, which is ${\cal O}(10\%)$ for the individual
   predictions and below $1\%$ for $\R138$. The envelope of the scale
   variations for $\R138$ shrinks in comparison with the envelopes for
   the individual predictions: from ${\cal O}(10\%)$ for the
   individual predictions to below $2\%$ for $\R138$ across most of
   the range in $\etg$.
\item In the incoherent case, the envelope of the scale variations for
  $\R138$ is ${\cal O}(14\%)$ in all regions of phase space.
\end{itemize}
A second approach is also investigated, which is free from ambiguity
in the correlation. It consists of considering the difference between the
LO and NLO predictions for $\R138$. The LO predictions are
obtained with \jetp\ using the same parameter settings and PDF set as
the baseline NLO predictions. The LO and NLO predictions for $\R138$
are compared and the differences are up to $3.5\%$, which are similar
to the estimates based on the standard approach with coherent
variations at the two centre-of-mass energies. Thus, the results of
this second approach support the use of the standard approach with a
coherent variation of the scales; an incoherent variation of the
scales clearly leads to an overestimation of the theoretical
uncertainty.

\begin{figure}[h]
\setlength{\unitlength}{1.0cm}
\begin{picture} (18.0,16.0)
\put (0.0,8.0){\includegraphics[width=8cm]{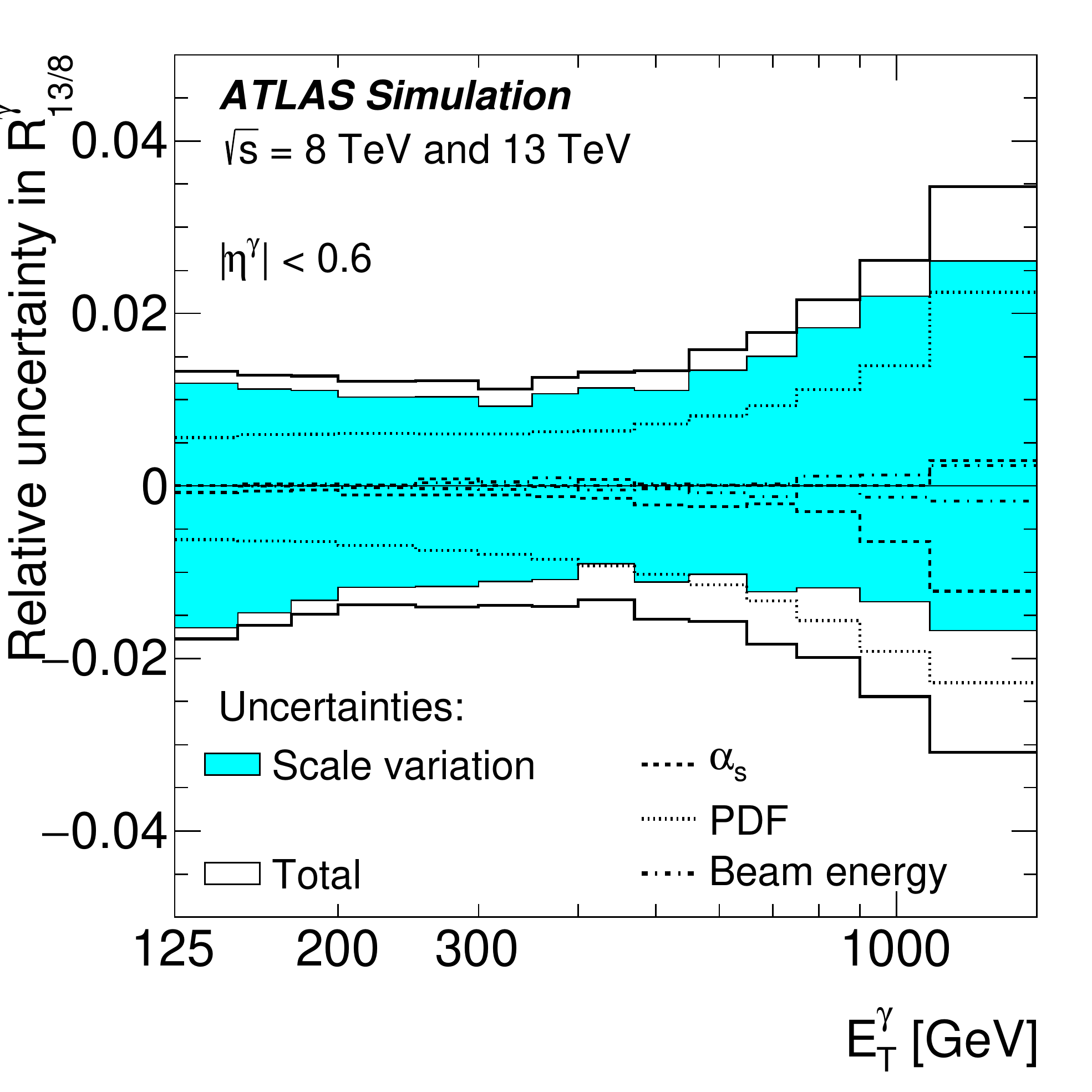}}
\put (8.0,8.0){\includegraphics[width=8cm]{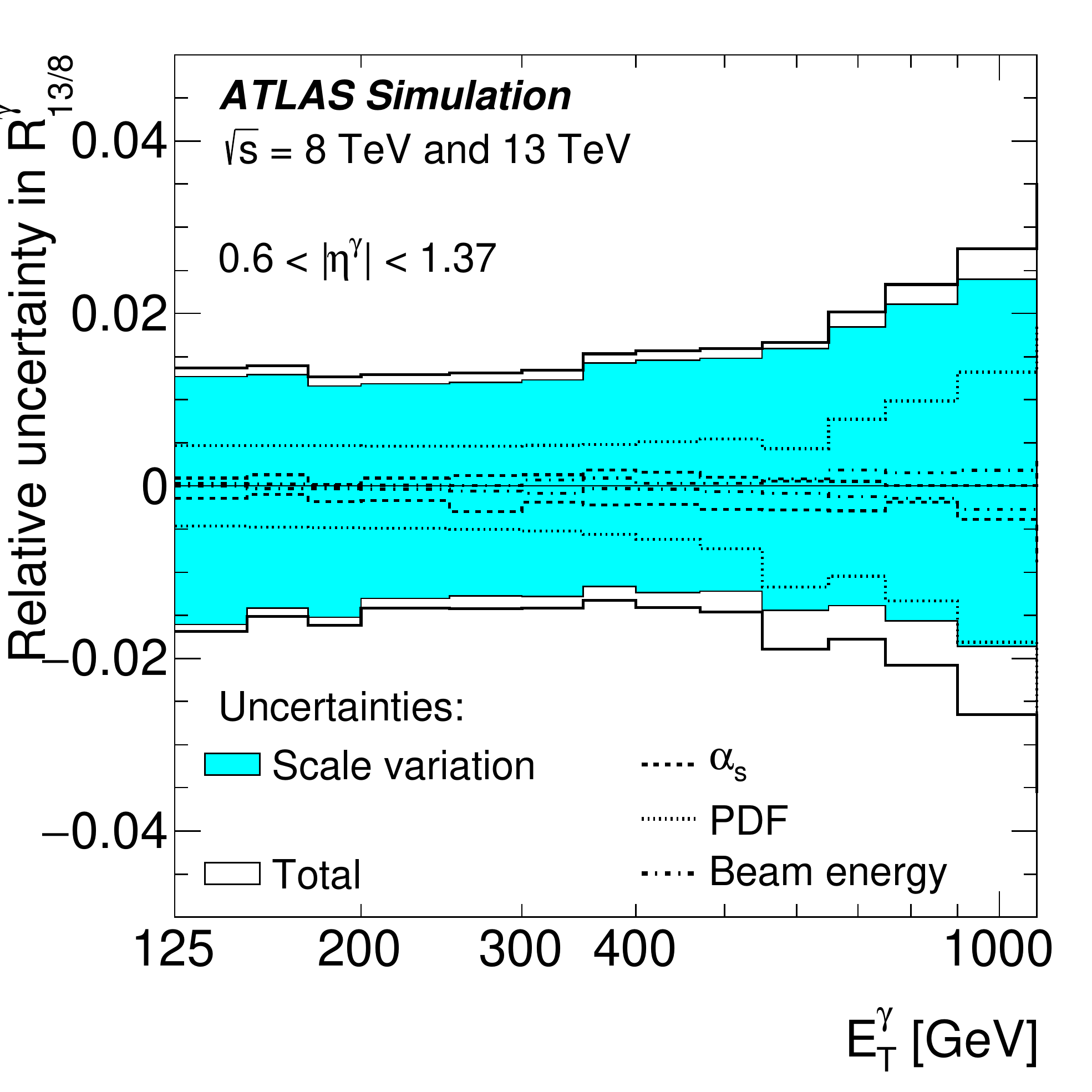}}
\put (0.0,0.0){\includegraphics[width=8cm]{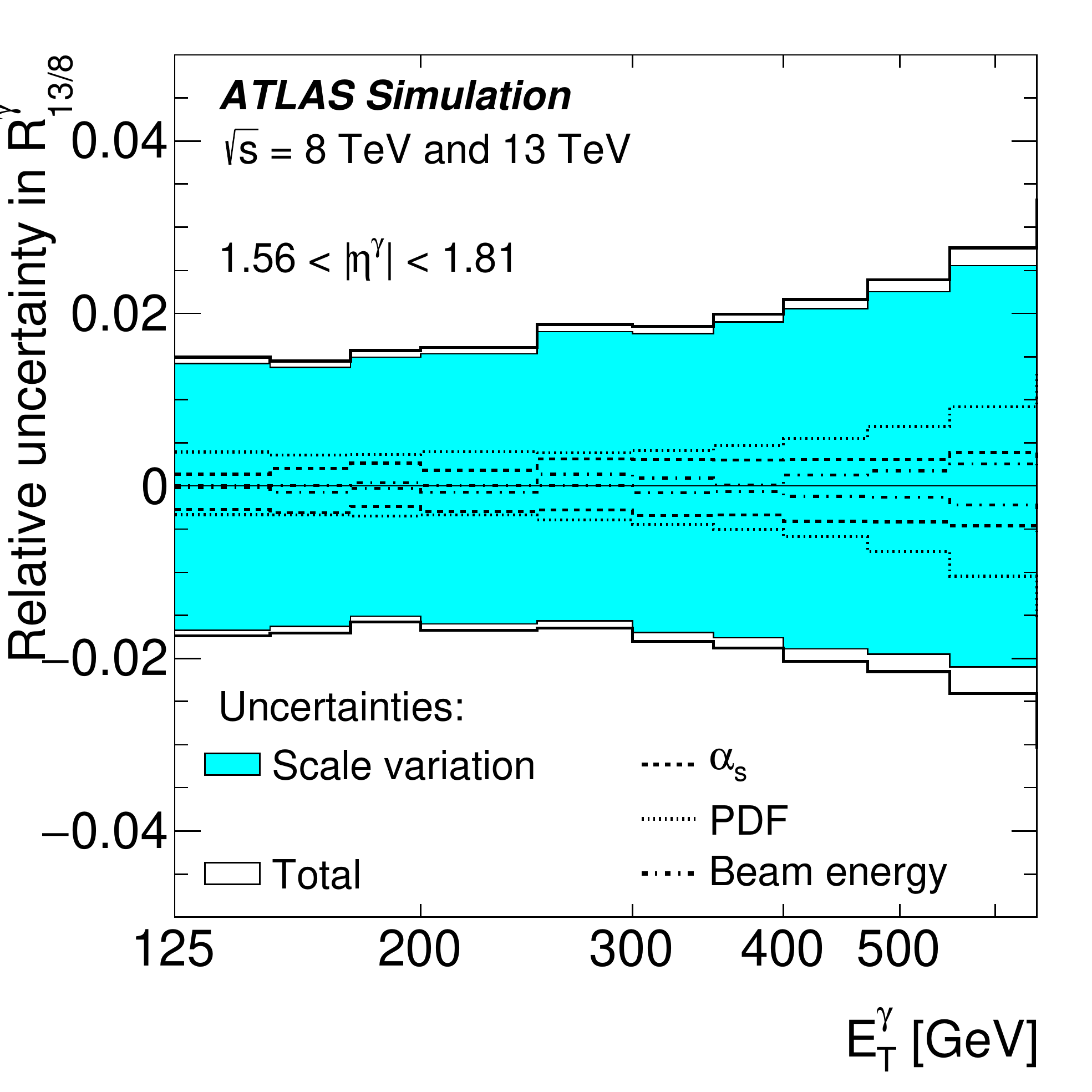}}
\put (8.0,0.0){\includegraphics[width=8cm]{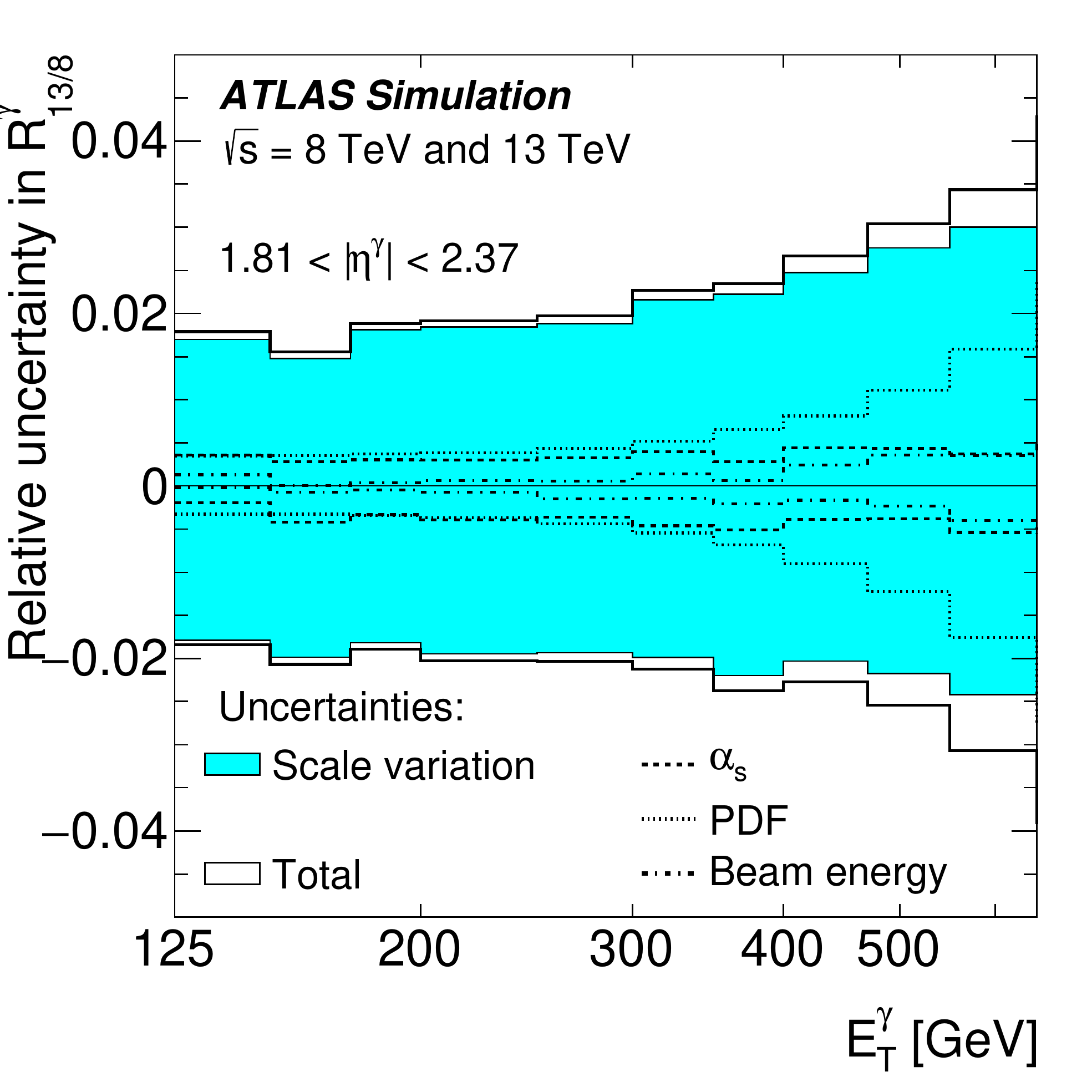}}
\end{picture}
\caption
{
  Relative theoretical uncertainty in $\R138$ as a function of $\etg$
  for different $\etag$ regions arising from the scale variations
  (shaded area), the value of $\as$ (dashed lines), the PDF (dotted
  lines) and the beam energy (dot-dashed lines). The total theoretical
  uncertainty is shown as the solid line.
}
\label{fig05}
\end{figure}

Figure~\ref{fig05} shows an overview of the theoretical uncertainties
in $\R138$. The total relative uncertainty is below $2\%$ ($4\%$) at
low (high) $\etg$ in all regions of $|\etag|$. The uncertainty due to
the variation of the scales is dominant everywhere. At high $\etg$ for
$|\etag|<0.6$ and $0.6<|\etag|<1.37$, the uncertainty due to the PDFs
can be as large as the contribution from the scale variations.

The NLO pQCD predictions of \jetp\ for $\R138$ based on the MMHT2014
parameterisations of the proton PDFs are about $2$ at
$\etg=125$~\GeV\ and increase as $\etg$ increases, to about $10$
for $\etg=1300\ (1000)$~\GeV\ for $|\etag|<0.6$
($0.6<|\etag|<1.37$). For $1.56<|\etag|<1.81$ ($1.81<|\etag|<2.37$),
the predicted $\R138$ increases from about $2$ at $\etg=125$~\GeV\ to
around $10$ ($25$) at $\etg=600$~\GeV. The increase is greater for the
forward regions than for the central regions. Predictions based on
different parameterisations of the proton PDFs are compared. Those
based on MMHT2014, NNPDF3.0 and CT14 are found to be similar in all
$\etag$ and $\etg$ regions. The predictions of $\R138$ based on
HERAPDF2.0 and ABMP16 show some differences from the
predictions based on the other PDFs in some regions of phase space,
especially at high $\etg$ (see Section~\ref{results}).

\subsection{Theoretical predictions for $\D138$}
\label{nlozratio}
The theoretical predictions for $\D138$ presented here are based on
NNLO QCD calculations for the predictions of $\Z138$ computed using
the program \dytub, which is an optimised version of the \dynn\
program~\cite{prl:98:222002,prl:103:082001},  and NLO QCD calculations
for the predictions of $\R138$ using \jetp\ with the procedure
described in Section~\ref{nlor}.

The calculations using \dytub\ are based on sets of PDFs extracted
using NNLO QCD fits, namely MMHT2014nnlo, CT14nnlo, HERAPDF2.0nnlo and
NNPDF3.0nnlo. The strong coupling constant $\asz$ is set to the value
assumed in the fit to determine the PDFs. In the case of MMHT2014nnlo
PDFs, $\asz$ is set to the value $0.118$.

For consistency, and to properly take into account the correlations in
the PDF uncertainties, the calculations of \jetp\ for $\R138$ are
repeated using the NNLO PDF sets mentioned above. It is consistent to
use NLO matrix elements convolved with PDF sets determined at
NNLO. The resulting predictions include partially NNLO corrections
and, therefore, are understood to still have NLO accuracy. For these
additional calculations, the same parameter settings for the number of
flavours, scales and fragmentation functions mentioned in
Section~\ref{nlor} are used. The change in the predictions for $\R138$
based on MMHT2014nnlo relative to those using MMHT2014nlo is $\sim
0.5\%$ at low $\etg$. At high $\etg$ the change depends on the
$|\etag|$ region: for $|\etag|<0.6$ the change is below $2\%$ for
$\etg<750$~\GeV\ and increases to $6\%$ in the highest-$\etg$ measured
point; for $0.6<|\etag|<1.37$ ($1.56<|\etag|<1.81$) the change is
below $2\%$ ($1.3\%$) for the entire measured range; for
$1.81<|\etag|<2.37$ the change is below $2\%$ for $\etg<550$~\GeV\ and
increases to $2.7\%$ in the highest-$\etg$ measured point.

The sources of uncertainty in the theoretical predictions based on
MMHT2014nnlo are the same as those described in
Section~\ref{nlor}. The uncertainty related to the beam energy is
neglected, due to the small size of its effect on $\R138$. The
uncertainties in the prediction of $\Z138$ due to the scale
variations, the PDFs and $\asz$ are $^{+0.02}_{-0.3}\%$,
$^{+0.9}_{-0.8}\%$ and $^{-0.03}_{-0.3}\%$, respectively. For the
predictions of $\D138$, the uncertainties have been estimated as
follows:

\begin{itemize}
\item The scale variations are considered uncorrelated between $Z$
  boson production and isolated-photon production since they are
  different processes.
\item The PDF uncertainties are considered fully correlated between
  $Z$ boson production and isolated-photon production.
\item The $\asz$ uncertainties are considered fully correlated between
  $Z$ boson production and isolated-photon production. The uncertainty
  in the predictions due to that in $\asz$ is estimated by using PDF
  sets in which $\asz$ was fixed at $0.116$ or $0.120$. 
\end{itemize}

In what follows, the resulting uncertainties in the predictions of
$\D138$ are described. In the region $|\etag|<0.6$
($0.6<|\etag|<1.37$), the total relative uncertainty is below $2\%$
for $125\leq\etg\leq 650\ (650)$~\GeV\ and it rises to $\approx
4.5\%\ (3.3\%)$ for $\etg=1300\ (1000)$~\GeV. In both $\etag$ regions,
the total uncertainty is mostly dominated by the variation of the
scales. For $|\etag|<0.6$ and $\etg\gtrsim 300$~\GeV, the
uncertainties in the PDFs are dominant, and for $0.6<|\etag|<1.37$ and
$\etg\gtrsim 750$~\GeV, the contributions from the scale variations
and the PDFs are equally large. In the region $1.56<|\etag|<1.81$
($1.81<|\etag|<2.37$), the total relative uncertainty is below $2\%\
(3\%)$ for $125\leq\etg\leq 350\ (470)$~\GeV\ and it rises to
$\approx 3\%\ (3.6\%)$ for $\etg=600$~\GeV. For $1.56<|\etag|<1.81$, the
uncertainty due to the variation of the scales is dominant, but for
$1.81<|\etag|<2.37$ and $\etg\gtrsim 550$~\GeV, the contributions from
the scale variations and the PDFs are equally important.

The theoretical predictions based on the MMHT2014nnlo
parameterisations of the proton PDFs for $\D138$ are about $1.4$ at
$\etg=125$~\GeV\ and increase as $\etg$ increases, to $6$--$17$ at
the high end of the spectrum, depending on the $\etag$ region. The
increase is larger for the forward regions than for the central
regions. Predictions based on different parameterisations of the
proton PDFs are compared; those based on MMHT2014nnlo, NNPDF3.0nnlo
and CT14nnlo are found to be similar in all $\etag$ and $\etg$
regions. The predictions of $\D138$ based on HERAPDF2.0nnlo show some
differences from the predictions based on the other PDFs in
some regions of phase space, especially at high $\etg$ (see
Section~\ref{results}).

\section{Experimental uncertainties}
\label{syst}

The sources of systematic uncertainties that affect the measurements
of the photon differential cross sections at $\sqrt s=8$ and $13$~\TeV\ are
detailed in Refs.~\cite{jhep:1608:005} and \cite{pl:b770:473},
respectively. A proper estimation of the systematic uncertainties in
this measurement of cross-section ratios requires taking into account
inter-$\!\sqrt s$ correlations for each source of systematic
uncertainty. Assuming no correlation provides a conservative
estimate and full correlation is used only when justified. The
estimation of the systematic uncertainties in the ratio has to take
into account the changes in the data-taking conditions as well as changes in
the detector conditions. The measurements at $\sqrt s =
8$~($13$)~\TeV\ are based on data taken when the LHC operated with a
bunch spacing of $50$~($25$)~ns. During the data-taking period at
$\sqrt s = 8$~($13$)~\TeV\ there were on average $20.7$ ($13.5$)
proton--proton interactions per bunch crossing. Furthermore, the
addition of the silicon-pixel insertable B-layer leads to extra
material upstream of the calorimeters for data-taking at $\sqrt s =
13$~\TeV. The procedures used to account for the impact of each source
of systematic uncertainty on the ratio $\R138$ are described below.

\subsection{Photon energy scale}
\label{geser}
The systematic uncertainties associated with the photon energy scale
and resolution represent the dominant experimental uncertainties in the
measurements of the differential cross sections for inclusive
isolated-photon production at both centre-of-mass energies. The
uncertainty arising from the photon energy scale ($\gamma$ES) in
$\R138$ is estimated by decomposing it into uncorrelated
sources for both the $8$~\TeV\ and $13$~\TeV\ measurements. A total of
$22$ individual components~\cite{epj:c74:3071} influencing the energy
scale and resolution of the photon are considered. Twenty of these
components are common to both centre-of-mass energies. 
For some of the components the uncertainty is separated into a part which
is correlated between the two centre-of-mass energies and another part
which is specific to $13$~\TeV\ data and which is treated as uncorrelated
(see below). These
components include the uncertainties in: the overall energy scale
adjustment using $Z\rightarrow\ele$ events; the non-linearity of the
energy measurement at the cell level; the relative calibration of the
different calorimeter layers; the amount of material in front of the
calorimeter; the modelling of the reconstruction of photon
conversions; the modelling of the lateral shower shape; the modelling
of the sampling term;\footnote{The relative energy resolution is
  parameterised as $\sigma(E)/E = a/\sqrt{E} \oplus c$, where $a$ is
  the sampling term and $c$ is the constant term.} and the measurement
of the constant term in $Z$ boson decays. The uncertainties depend on
$\etg$ as well as on $|\etag|$ and are larger in the region
$1.56<|\etag|<1.81$ due to the presence of more material
than in other $|\etag|$ regions. The remaining two
components are specific to the $13$~\TeV\ measurement and take into
account the differences in the configuration of the ATLAS detector
between 2012 and 2015, namely  changes in the LAr temperature, in the
stability of the layer intercalibration and in the
material in front of the calorimeters between Run 1 and Run
2~\cite{ATL-PHYS-PUB-2016-015}.

The procedure used to estimate the systematic uncertainty in $\R138$
is as follows: all the uncertainty components described above are
taken as fully correlated except for the uncertainty in the overall
energy scale adjustment using $Z\rightarrow\ele$ events, which for 2015
includes the effects of the changes in the configuration of the ATLAS
detector mentioned above, and the uncertainties specific to the
$13$~\TeV\ measurement. Calibration differences due to a change of 
optimal filtering coefficients and LAr timing samples between Run 1 and 
Run 2 are considered as a source of uncertainty in $\R138$.
The uncertainties in the photon energy scale due to pile-up are small enough
compared to other uncertainties that the specific treatment of the correlation
does not impact the results. The uncertainties due the photon energy resolution
are treated as uncorrelated between $\sqrt s= 13$~\TeV\ and $8$~\TeV\ since
they include the effects of pile-up, which was different in the 2012 and 2015
data-taking periods. 

\begin{figure}[h]
\setlength{\unitlength}{1.0cm}
\begin{picture} (18.0,15.6)
\put (0.0,7.8){\includegraphics[width=8cm]{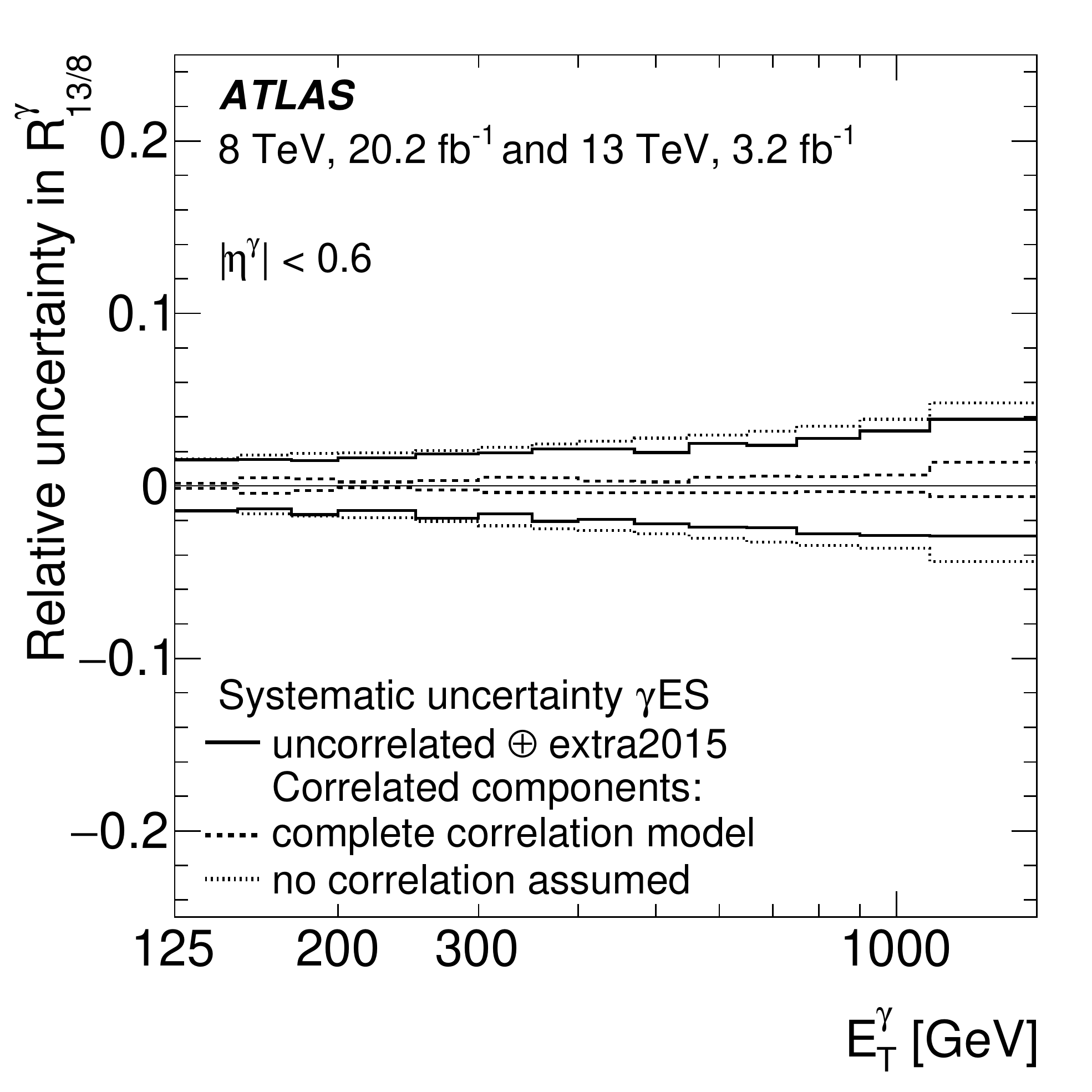}}
\put (8.0,7.8){\includegraphics[width=8cm]{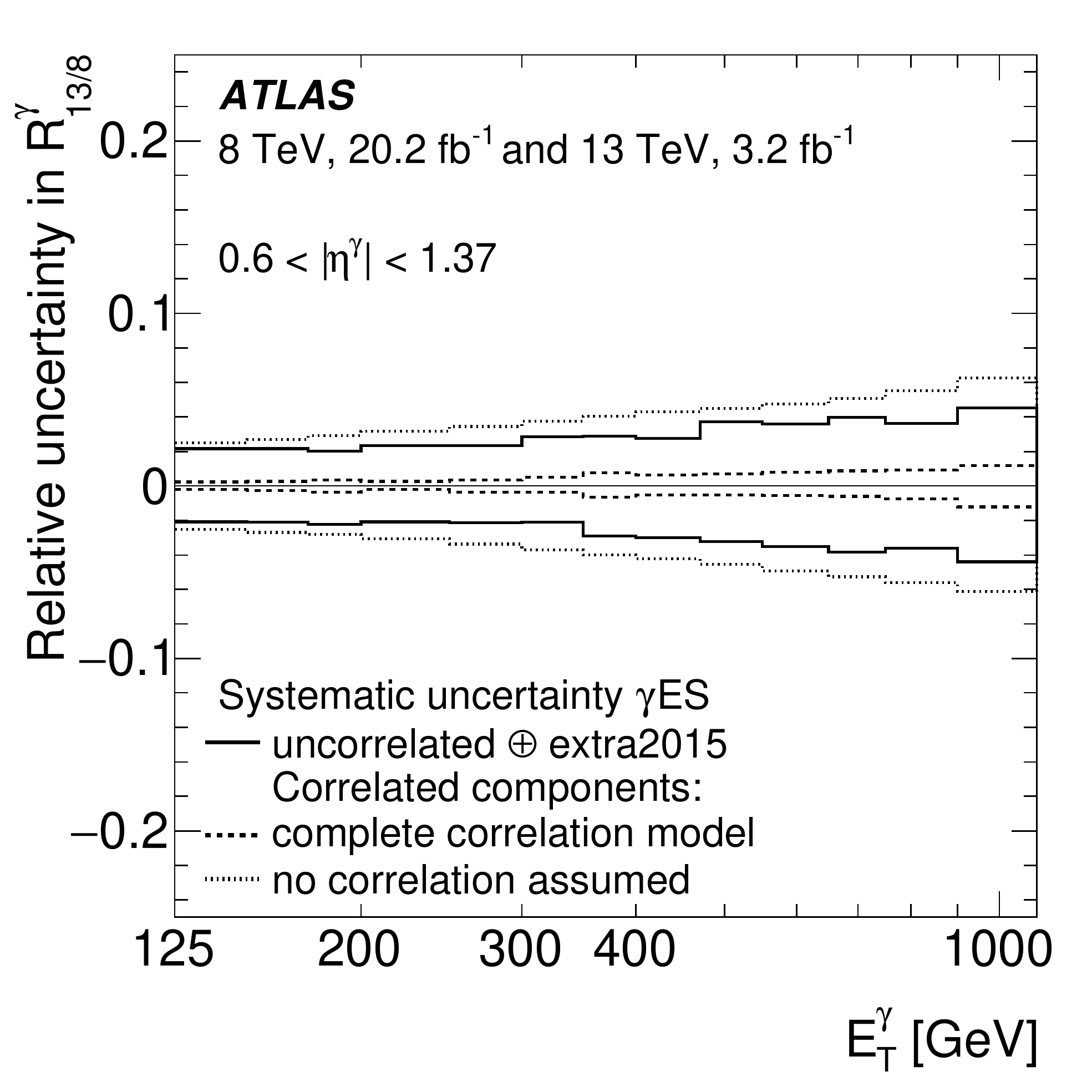}}
\put (0.0,0.0){\includegraphics[width=8cm]{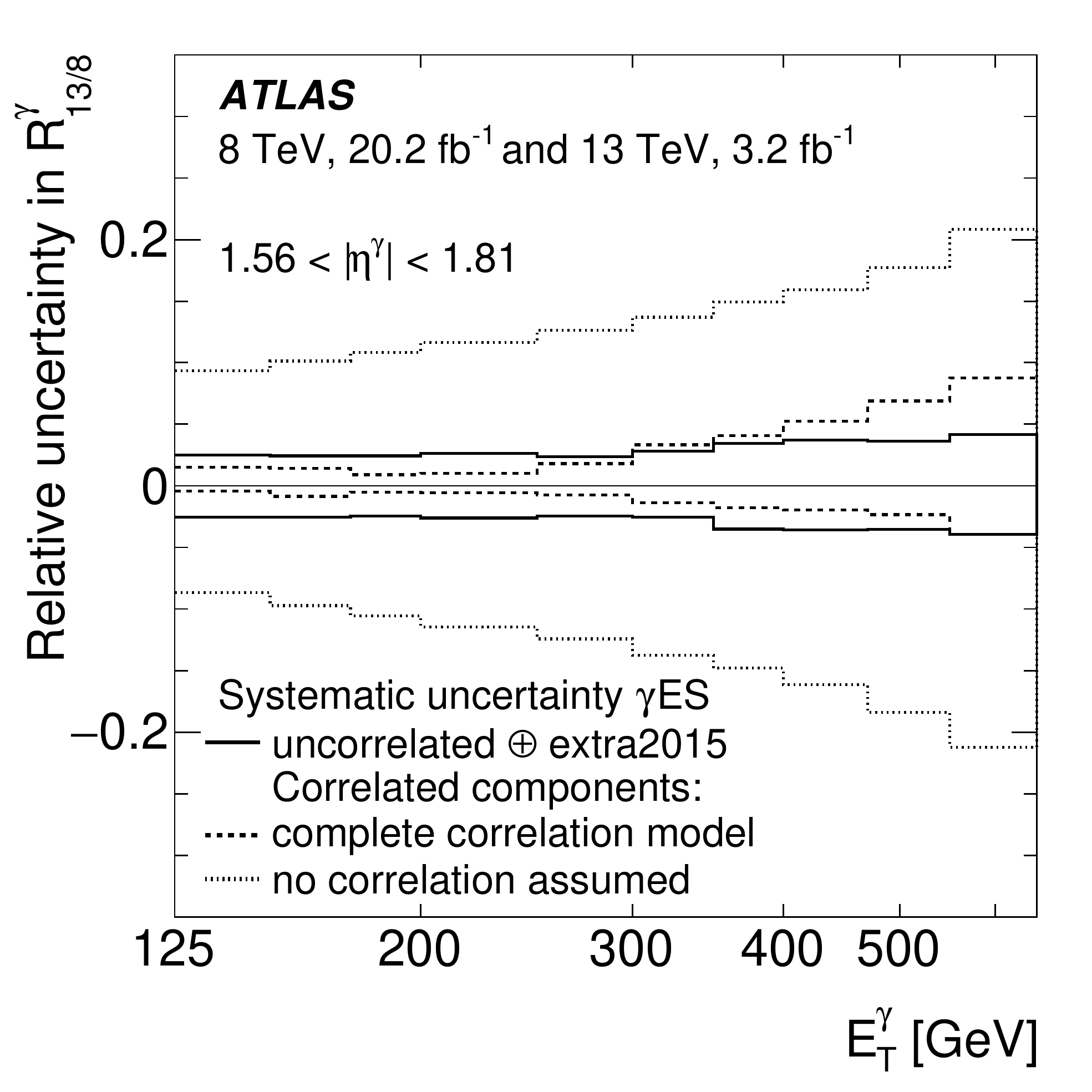}}
\put (8.0,0.0){\includegraphics[width=8cm]{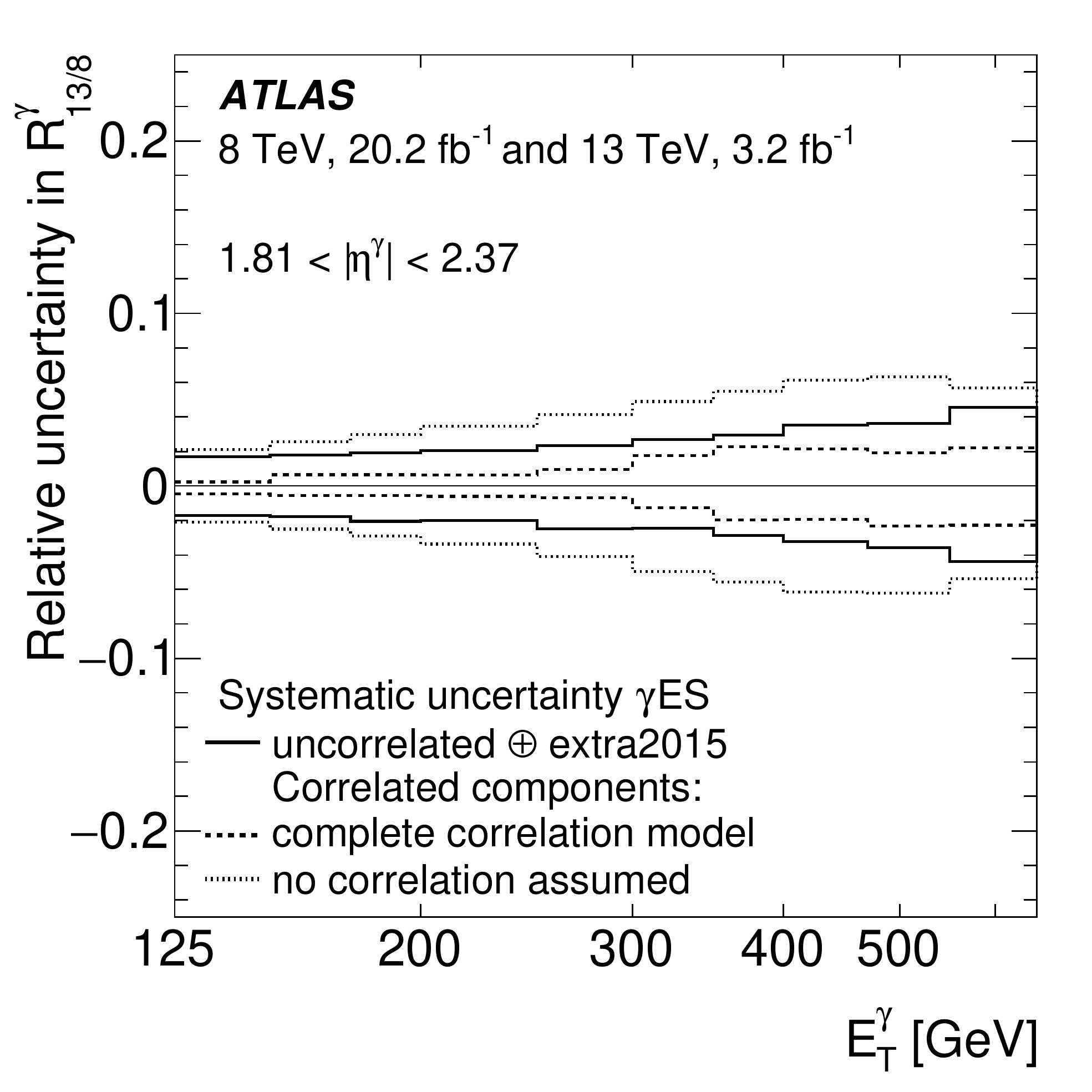}}
\end{picture}
\caption
{
  Relative systematic uncertainty in $\R138$ as a function of $\etg$
  for different $\etag$ regions due to the $\gamma$ES correlated
  components (dashed lines). For comparison, the results of
  considering the components as uncorrelated are also shown (dotted
  lines) to illustrate the reduction in the size of the systematic
  uncertainty when the proper treatment is applied. The relative
  uncertainty due to the uncorrelated components of the photon energy
  scale and the components specific to 2015 is also shown (solid
  lines).
}
\label{fig01}
\end{figure}

The relative uncertainty due to the correlated components of the
photon energy scale in $\R138$ as a function of $\etg$ is shown in
Figure~\ref{fig01} for each region in $\etag$. For illustration
purposes, the result of estimating that part of the systematic
uncertainty assuming no correlation is also shown in this figure: the
results obtained using the complete correlation model exhibit a large
reduction in comparison with those in which the correlations are
ignored. This demonstrates that a proper treatment of the
inter-$\!\sqrt s$ correlations in the systematic uncertainties
associated with the photon energy scale is important. In addition, the
relative uncertainty due to the uncorrelated components of the photon
energy scale and the components specific to 2015 is also shown in
Figure~\ref{fig01}.

\FloatBarrier

\subsection{Other sources of experimental uncertainty} 
\label{othersyst}
The other sources of experimental uncertainty affecting the
measurements are treated as listed below. For several of these
sources, the uncertainties in the measurements at $\sqrt s=13$~\TeV\
and $8$~\TeV\ are treated conservatively as uncorrelated since their
impact is small.
\begin{itemize}
\item {\textbf{Statistical uncertainties.}} The statistical
  uncertainties in both the data and the Monte Carlo simulations at $\sqrt
  s=13$ and $8$~\TeV\ are treated as uncorrelated.
\item {\textbf{Luminosity uncertainty.}} 
The luminosity uncertainties associated to the measurements of the
photon cross sections at $\sqrt s=8$~\TeV\ and $13$~\TeV\  are
dominated by effects that are uncorrelated between different
centre-of-mass energies and data-taking periods. The resulting relative
uncertainty in $\R138$ amounts to $\pm 2.8\%$.
\item {\textbf{Trigger uncertainty.}} The uncertainties in the trigger
  efficiency are treated as uncorrelated for data at different $\sqrt
  s$. Different trigger requirements were used during 2012 and
  2015. In addition, during 2012 a three-level trigger system was used
  to select events while in 2015 a two-level system was employed.
\item {\textbf{Photon-identification uncertainty.}} In both
  measurements, the photon identification is based primarily on shower
  shapes in the EM calorimeter. These uncertainties are treated as
  uncorrelated since different methods are used at $\sqrt
  s=13$~\TeV~\cite{ATL-PHYS-PUB-2016-014} and $8$~\TeV~\cite{epj:c76:666}
  to estimate the uncertainties; in addition, the photon
  identification criteria are re-optimised for data taken at
  $13$~\TeV.
\item {\textbf{Modelling of the photon isolation in Monte Carlo.}}
  In both measurements, the photon candidate is required to be
  isolated. The in-time (out-of-time) pile-up, which is due to
  additional $pp$ collisions in the same (earlier or later than) bunch
  crossing as the event of interest, was different in 2012 and 2015
  due to the different LHC conditions, namely the instantaneous
  luminosity and the bunch spacing. For simulated events, data-driven
  corrections to $\etiso$ are applied such that the peak position in
  the $\etiso$ distribution coincides in data and simulation. These
  uncertainties are treated as uncorrelated since different methods
  are used at $\sqrt s=13$ and $8$~\TeV\ for the corrections and
  uncertainties.
\item {\textbf{Choice of background control regions.}} 
  The background subtraction is performed using a data-driven
  two-dimensional sideband technique based on background control
  regions. A plane is formed by the variable $\etiso$ and a binary
  variable that encapsulates the photon identification (``tight''
  vs. ``non-tight''). A photon candidate is classified as
  ``non-tight'' if it fails at least one of four requirements on the
  shower-shape variables computed from the energy deposits in the
  first layer of the EM calorimeter, but satisfies the tight
  requirement on the total lateral shower width in the first layer and
  all the other tight identification criteria in other
  layers~\cite{epj:c76:666}.  The plane is divided into four regions:
  region A for tight isolated photons, region B for tight non-isolated
  photons, region C for non-tight isolated photons and region D for
  non-tight non-isolated photons. The background control regions B, C
  and D are specified by lower and upper limits on $\etiso$ as well as
  by the definition of ``non-tight'' photon candidates. Variations of
  the limits and alternative definitions of  the ``non-tight''
  condition are used to estimate the uncertainties due to the choice
  of background control regions.  These uncertainties are treated as
  uncorrelated since, as mentioned above, the photon-identification
  requirements are re-optimised for data-taking at $13$~\TeV.
\item {\textbf{Photon identification and isolation correlation in the
      background.}} In the background subtraction method described
  above, the photon isolation and identification variables are assumed
  to be uncorrelated for background events. Uncertainties due to this
  assumption are estimated by using validation regions, which are
  dominated by background. These uncertainties are treated as
  uncorrelated since, as mentioned above, the photon-identification
  requirements are re-optimised for data-taking at $13$~\TeV.
\item {\textbf{Signal modelling.}} MC simulations of signal processes
  are used to estimate the signal leakage fractions in the background
  control regions and to compute the unfolding corrections. For both
  measurements, at $\sqrt s=13$ and $8$~\TeV, the
  \pyt~\cite{cpc:178:852} generator is used for the nominal results
  and the \sher~\cite{jhep:0902:007}  generator for studies of
  systematic uncertainties related to the model dependence. The
  uncertainty due to the mixture of direct and fragmentation processes
  in the simulations is estimated using the MC simulations of
  \pyt. These uncertainties are treated as uncorrelated since
  different methods and versions of the generators are used at $\sqrt
  s=13$~\TeV\ and $8$~\TeV\ to estimate the uncertainties. For $\sqrt
  s=13$~($8$)~\TeV, \pyt~8.186 with the A14 tune (\pyt~8.165 with the
  AU2 tune) and \sher~2.1.1 (\sher~1.4.0) with the CT10 tune are
  used. For the $8$~\TeV\ analysis the results of using the default
  admixture of direct and fragmentation contributions in \pyt\ are
  compared with those using an optimal admixture obtained by fitting the
  two components to the data; for the $13$~\TeV\ analysis the results
  of enhancing the fragmentation contribution by a factor of two or
  removing it completely are compared with those using the default
  admixture.
\item {\textbf{QCD-cascade and hadronisation model dependence.}} These
  uncertainties are treated as uncorrelated since different versions
  and tunes of the Monte Carlo generators \pyt\ and \sher\ are used at
  $\sqrt s=13$ and $8$~\TeV.
\item {\textbf{Pile-up uncertainties.}} The in-time and out-of-time
  pile-up in the 2012 and 2015 data-taking periods were
  different. Conservatively and given the fact that the impact is
  rather small, these uncertainties are treated as uncorrelated.
\end{itemize}

\subsection{Total experimental uncertainties in $\R138$} 
\label{expratio}
\begin{figure}[h]
\setlength{\unitlength}{1.0cm}
\begin{picture} (18.0,16.0)
\put (0.0,8.0){\includegraphics[width=8cm]{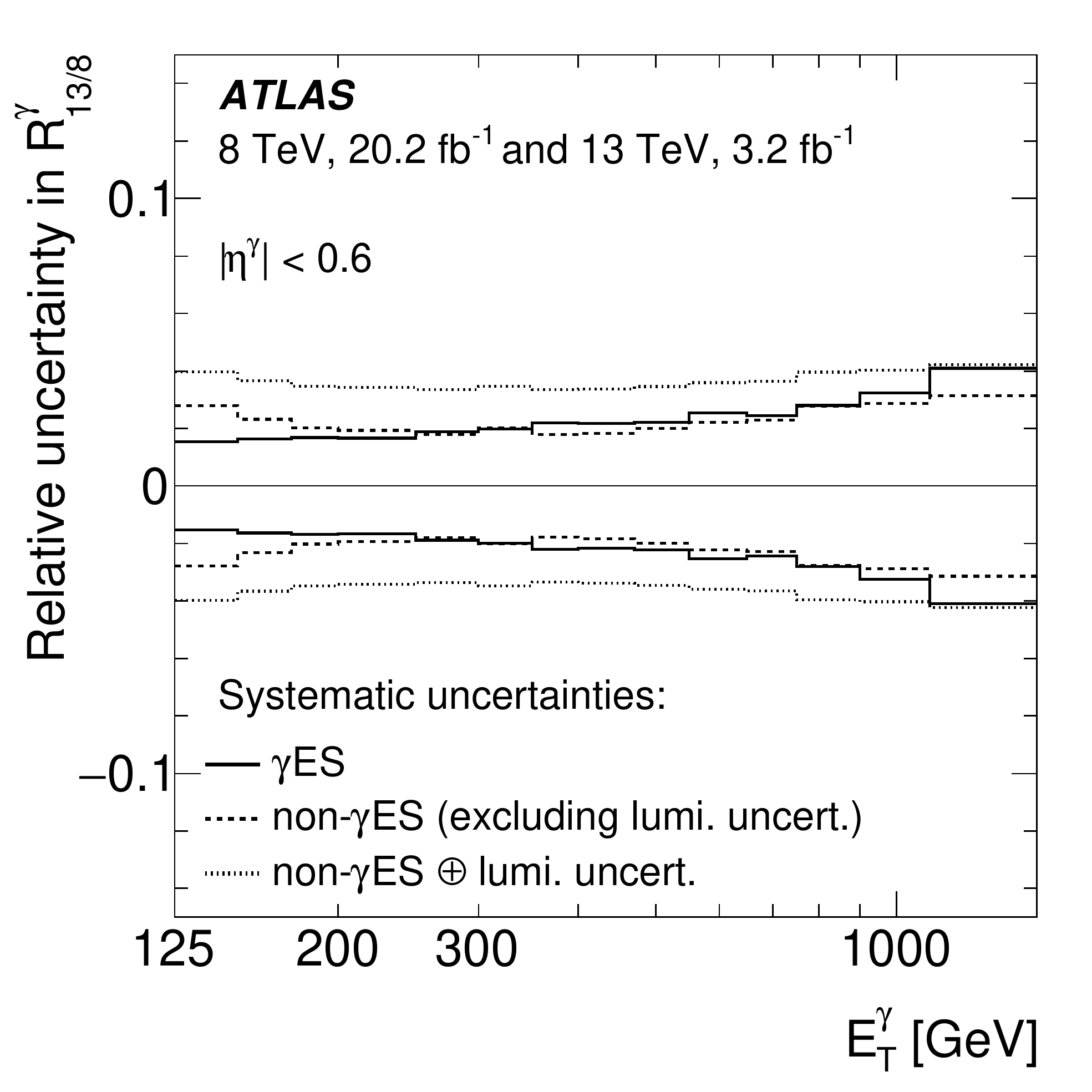}}
\put (8.0,8.0){\includegraphics[width=8cm]{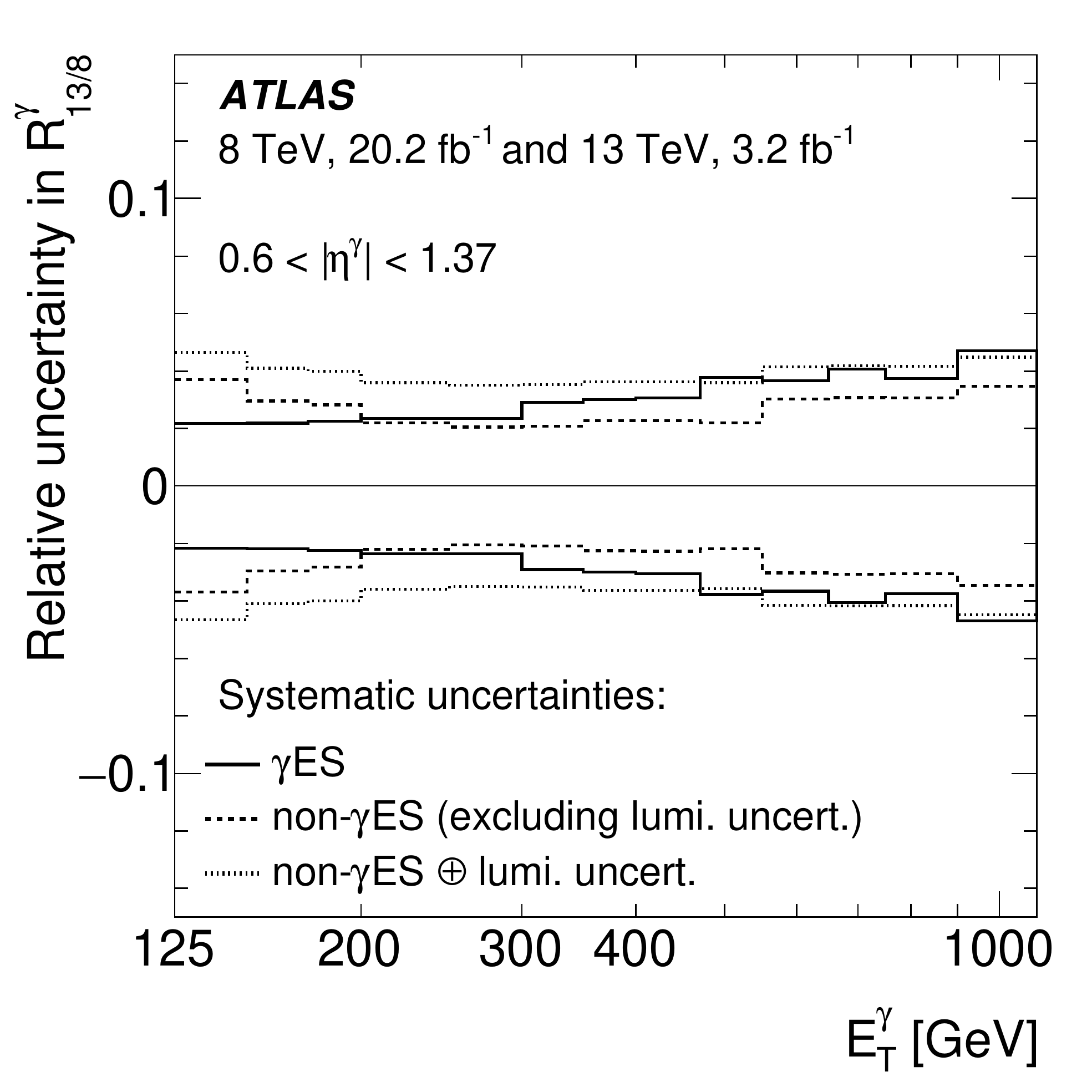}}
\put (0.0,0.0){\includegraphics[width=8cm]{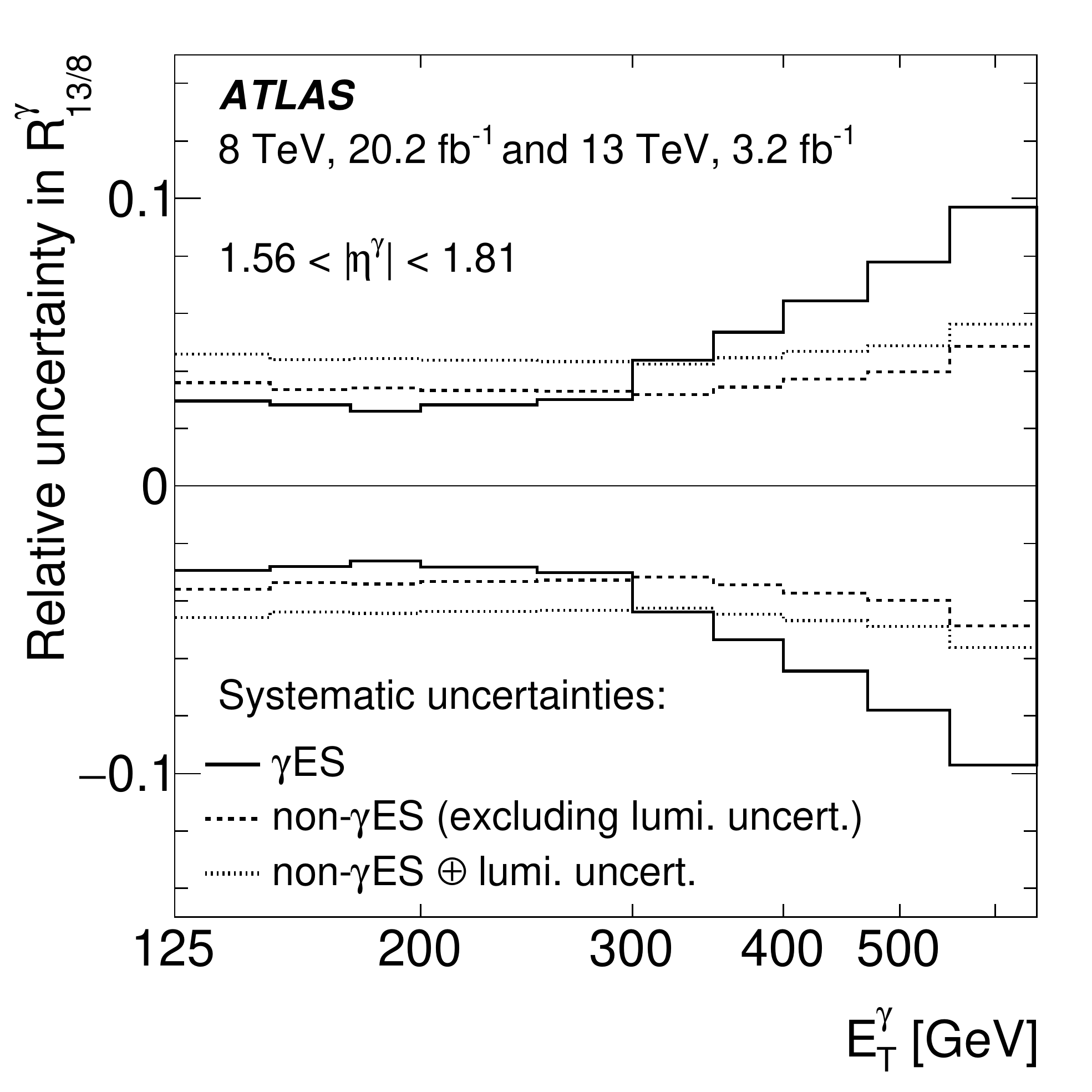}}
\put (8.0,0.0){\includegraphics[width=8cm]{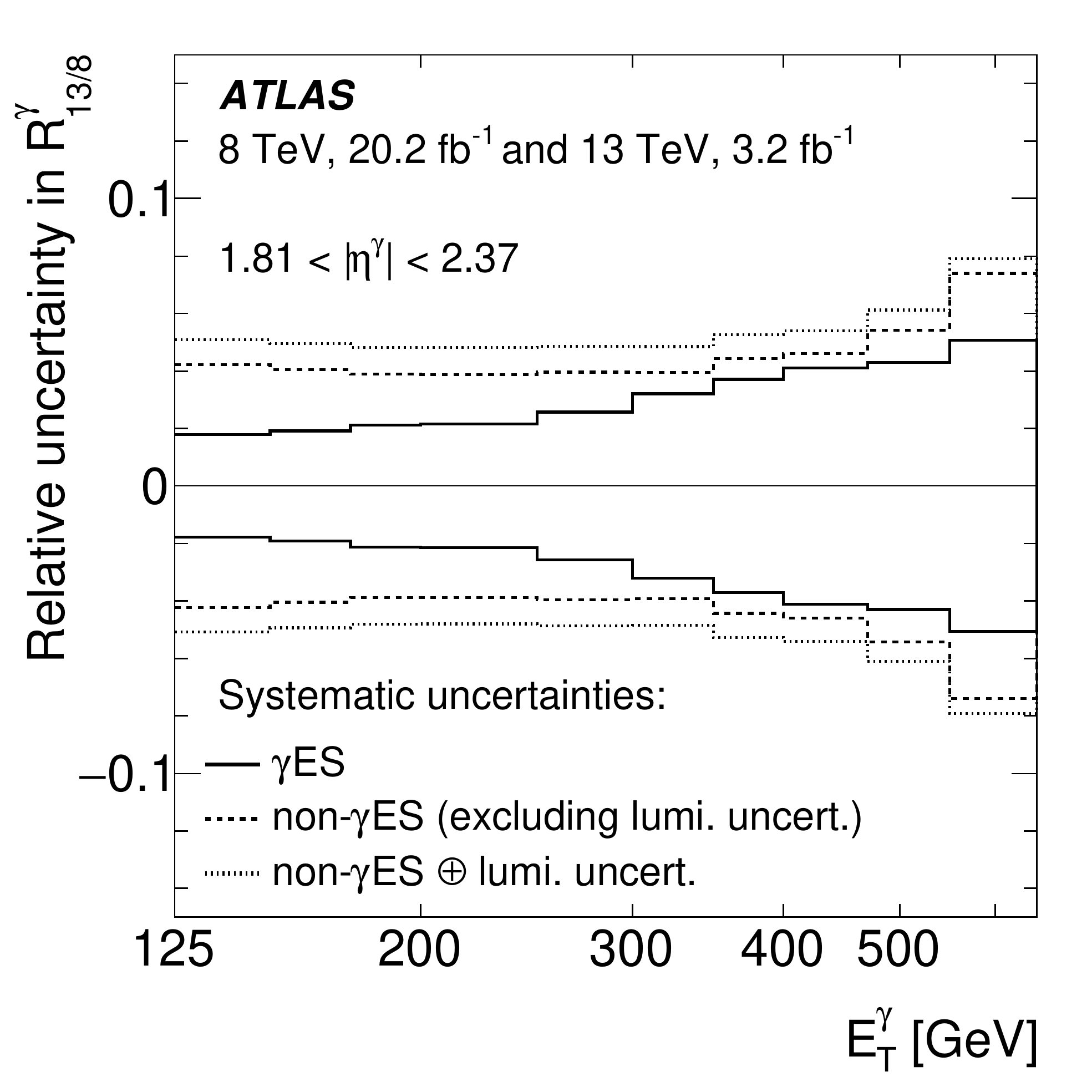}}
\end{picture}
\caption
{
  Relative systematic uncertainty in $\R138$ as a function of $\etg$
  for different $\etag$ regions due to different sources: $\gamma$ES
  uncertainties (solid lines), non-$\gamma$ES uncertainties excluding
  the luminosity uncertainty (dashed lines) and non-$\gamma$ES and
  luminosity uncertainties added in quadrature (dotted lines).
}
\label{fig02}
\end{figure}

Using the prescription for the treatment of the correlations between
the measurements described in the previous sections, the systematic
uncertainties in $\R138$ are evaluated. Figure~\ref{fig02} shows the
relative uncertainties in $\R138$ due to (i) the photon energy scale,
which includes the correlated and uncorrelated contributions as well
as the additional ones associated with 2015 data, (ii) the remaining
sources of systematic uncertainty excluding that in the luminosity
measurements and (iii) the sum in quadrature of the non-$\gamma$ES
uncertainties and the uncertainty due to the luminosity
determination. The uncertainty due to the photon energy scale
increases as $\etg$ increases and is larger for the region
$1.56<|\etag|<1.81$ due to more material in
front of the calorimeters than in the other regions. From
Figure~\ref{fig02} it is concluded that the relative uncertainty in
$\R138$ due to the photon energy scale is no longer the dominant
uncertainty, except for $\etg>300$~\GeV\ in the regions
$0.6<|\etag|<1.37$ and $1.56<|\etag|<1.81$.

\begin{figure}[h]
\setlength{\unitlength}{1.0cm}
\begin{picture} (18.0,16.0)
\put (0.0,8.0){\includegraphics[width=8cm]{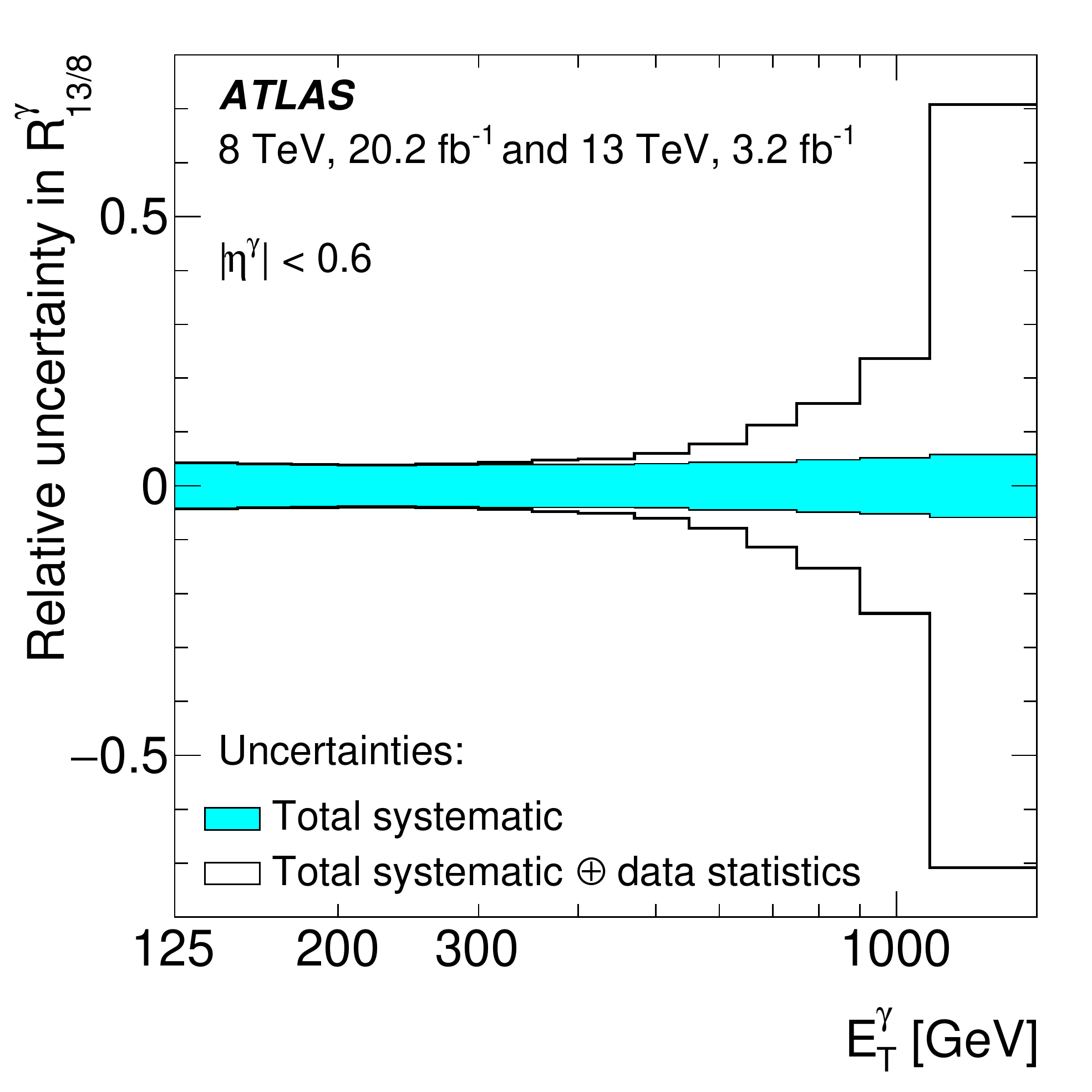}}
\put (8.0,8.0){\includegraphics[width=8cm]{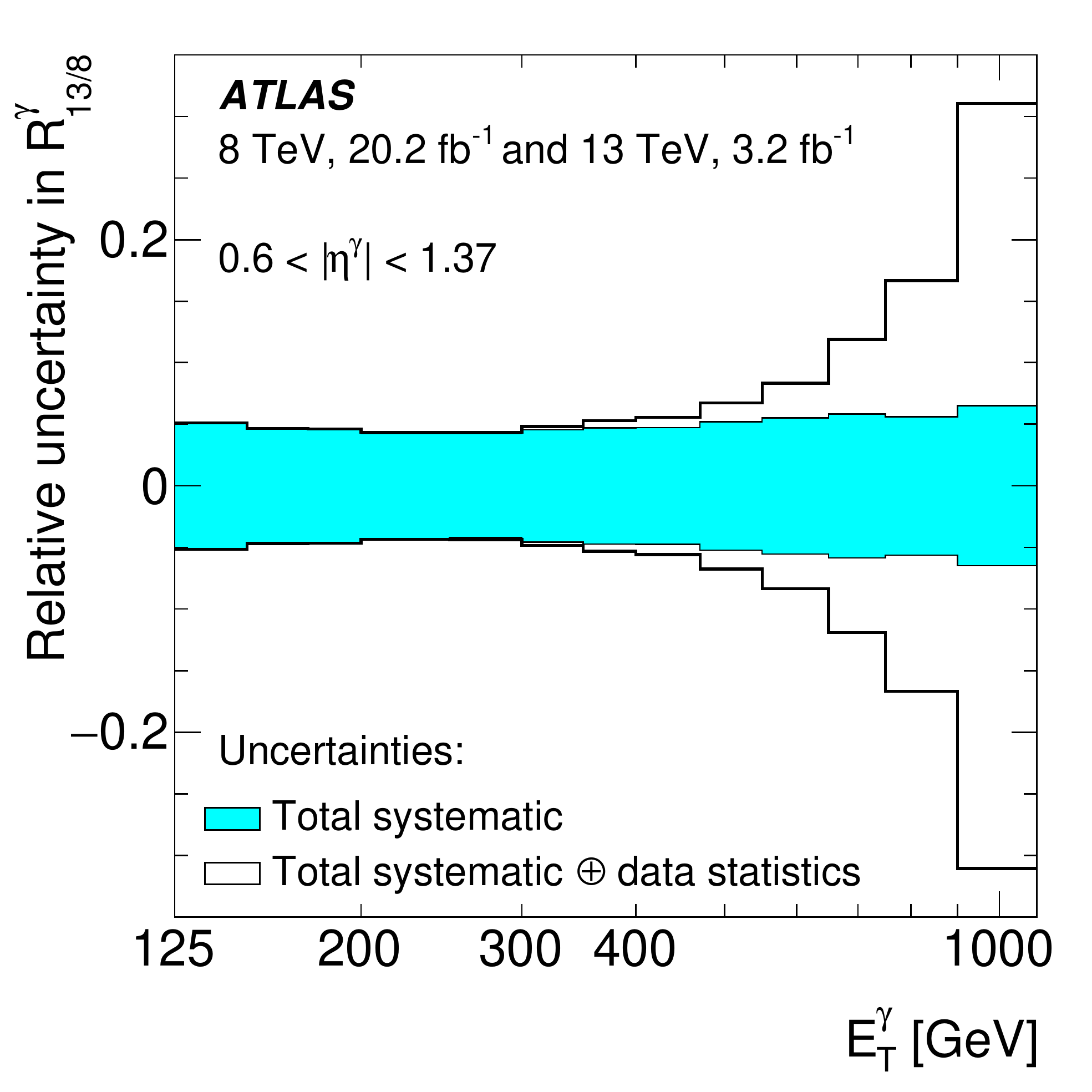}}
\put (0.0,0.0){\includegraphics[width=8cm]{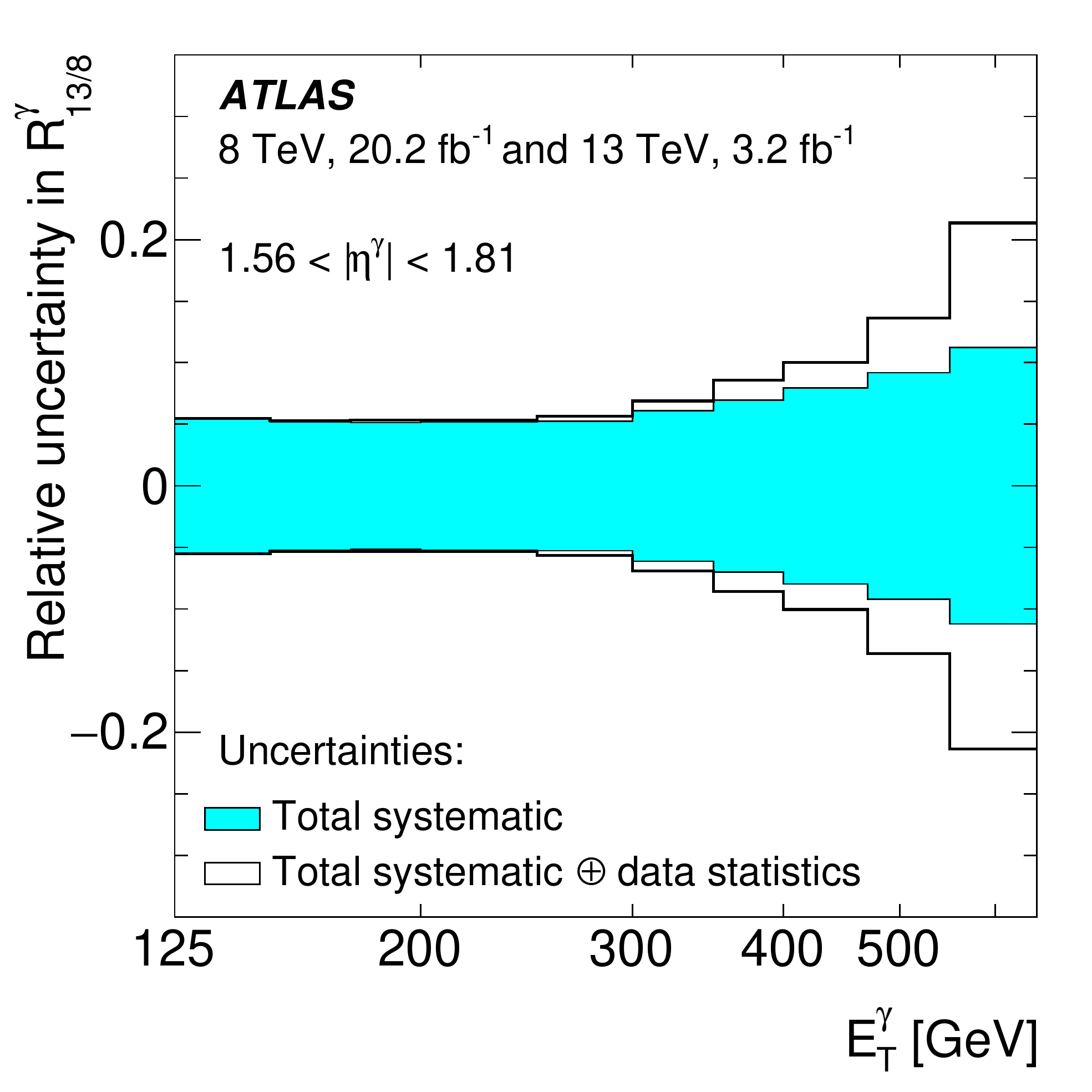}}
\put (8.0,0.0){\includegraphics[width=8cm]{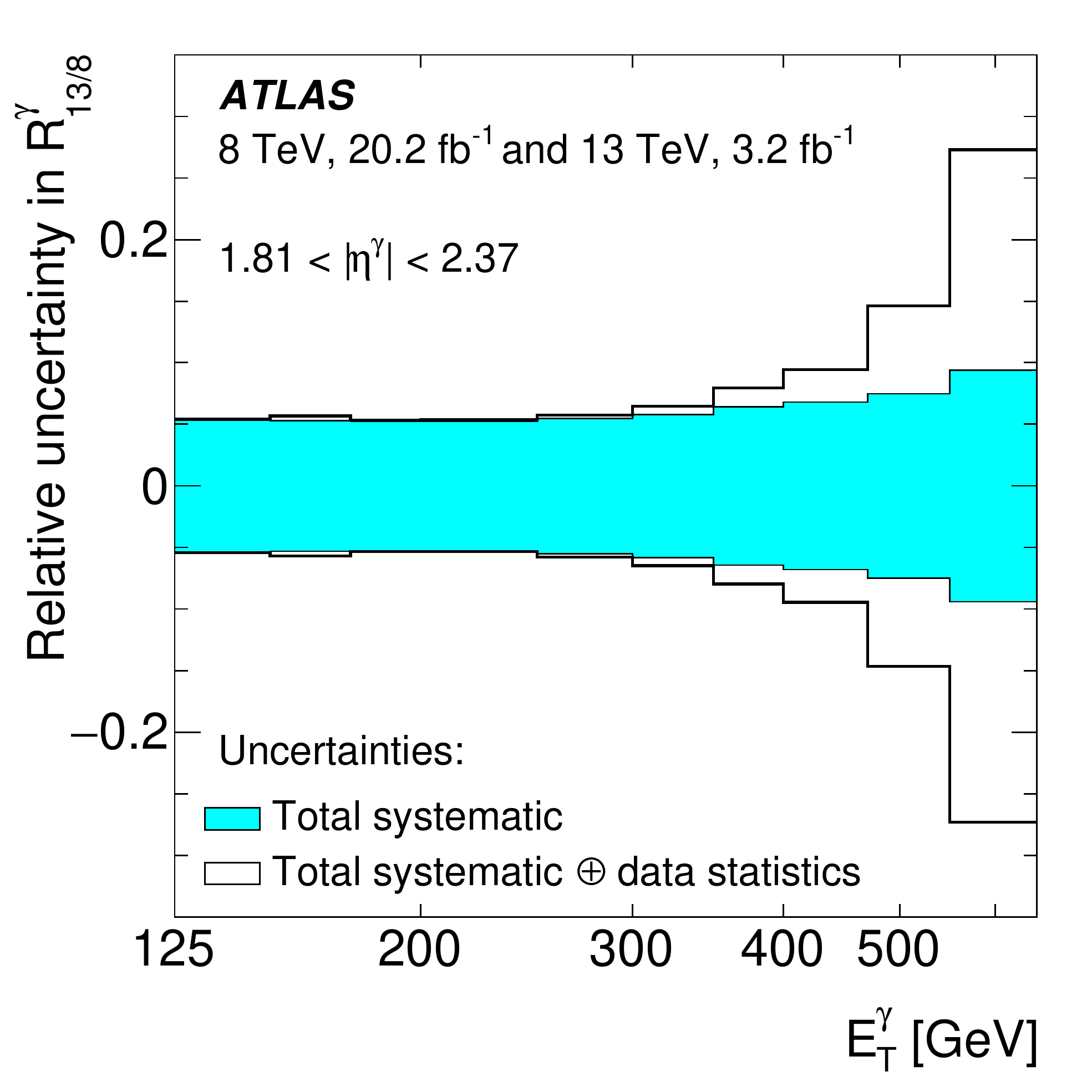}}
\end{picture}
\caption
{
  Total relative systematic uncertainty in $\R138$ as a function of
  $\etg$ for different $\etag$ regions (shaded band) and
  the sum in quadrature of the total relative systematic and
  statistical uncertainties  (solid line).
}
\label{fig03}
\end{figure}

The total relative experimental systematic uncertainty in $\R138$ is
shown in Figure~\ref{fig03}, as is its sum in quadrature with the
relative statistical uncertainty. In all pseudo-rapidity regions, the
systematic uncertainty is dominant compared to the statistical
uncertainty up to $\etg\sim 300$~\GeV,
while the measurement becomes statistically
limited for $\etg \gtrsim 600$~\GeV. There are significant
correlations in the systematic uncertainties across bins in $\etg$;
the uncertainty in the luminosity measurement is one of the
major contributions and is fully correlated for all bins in $\etg$ and all
$\etag$ regions.

\subsection{Total experimental uncertainties in $\D138$} 
\label{expzratio}
The total relative experimental uncertainty in $\D138$ is obtained as
follows:
\begin{itemize}
\item 
  The uncertainty in $\R138$ as presented in Section~\ref{othersyst},
  not including the contribution from the luminosity, is used. The
  uncertainty in the luminosity measurement cancels out in $\D138$
  since the measurements of $\R138$ and $\Z138$ are performed using
  data taken during the same periods of 2012 and 2015.
\item The statistical ($0.1\%$) and systematic ($0.7\%$) uncertainties
  in $\Z138$ are added in quadrature to the total uncertainty in
  $\R138$ (see Section~\ref{expratio}). The systematic uncertainty in
  $\Z138$ is dominated by the uncertainty in the lepton reconstruction
  and efficiency; the correlation between the small contribution due
  to the electron energy scale and the photon energy scale in $\D138$
  can be safely neglected.
\end{itemize}

\begin{figure}[h]
\setlength{\unitlength}{1.0cm}
\begin{picture} (18.0,16.0)
\put (0.0,8.0){\includegraphics[width=8cm]{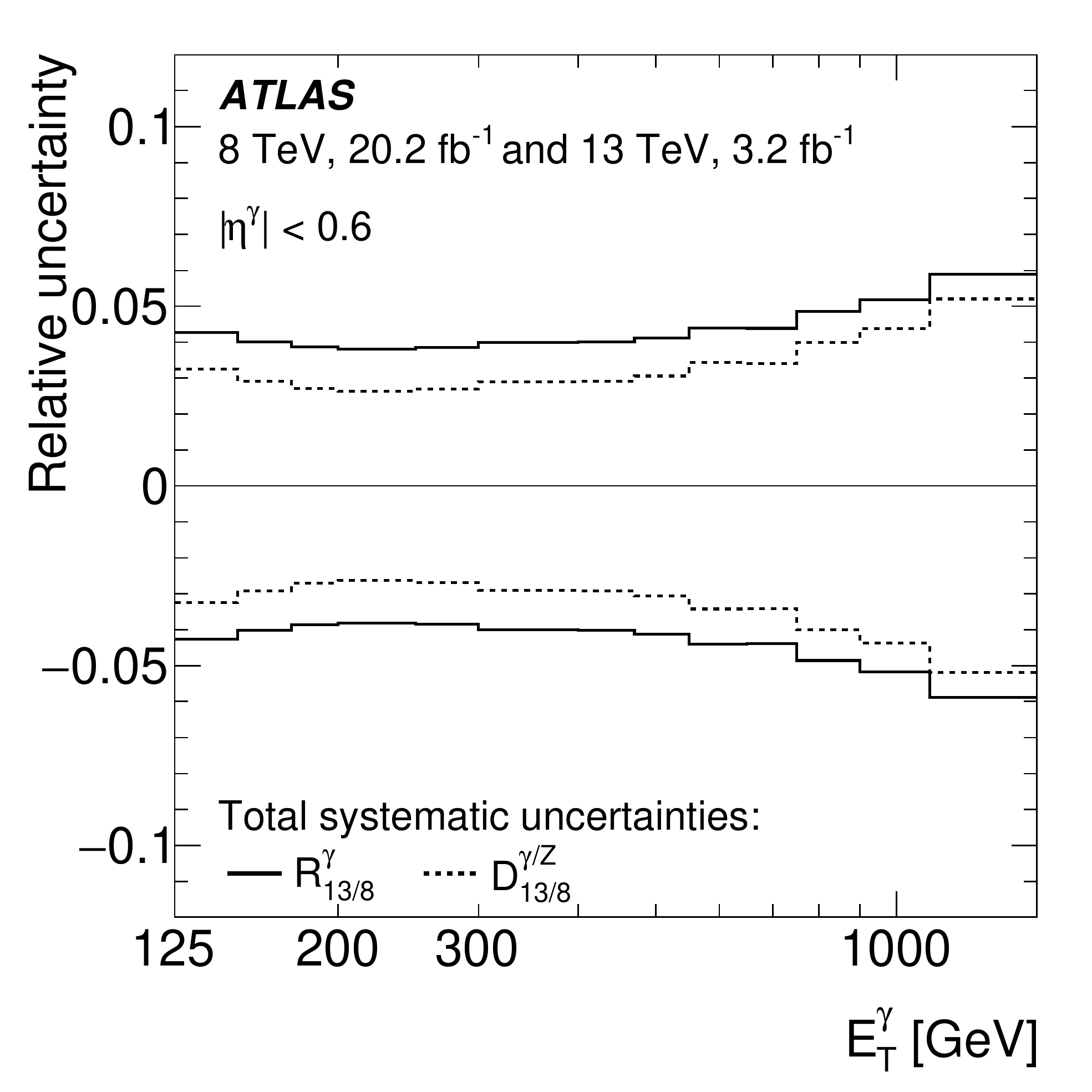}}
\put (8.0,8.0){\includegraphics[width=8cm]{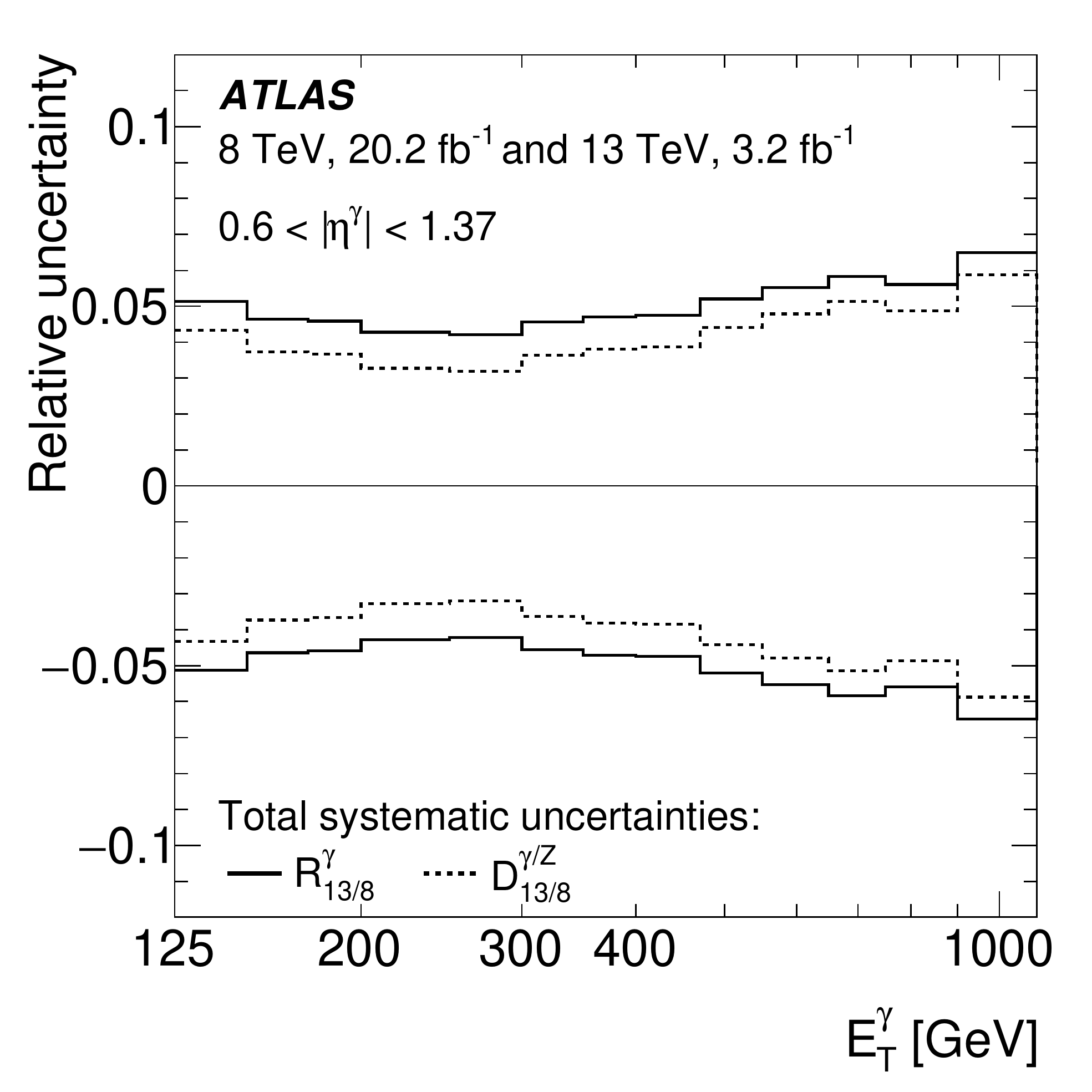}}
\put (0.0,0.0){\includegraphics[width=8cm]{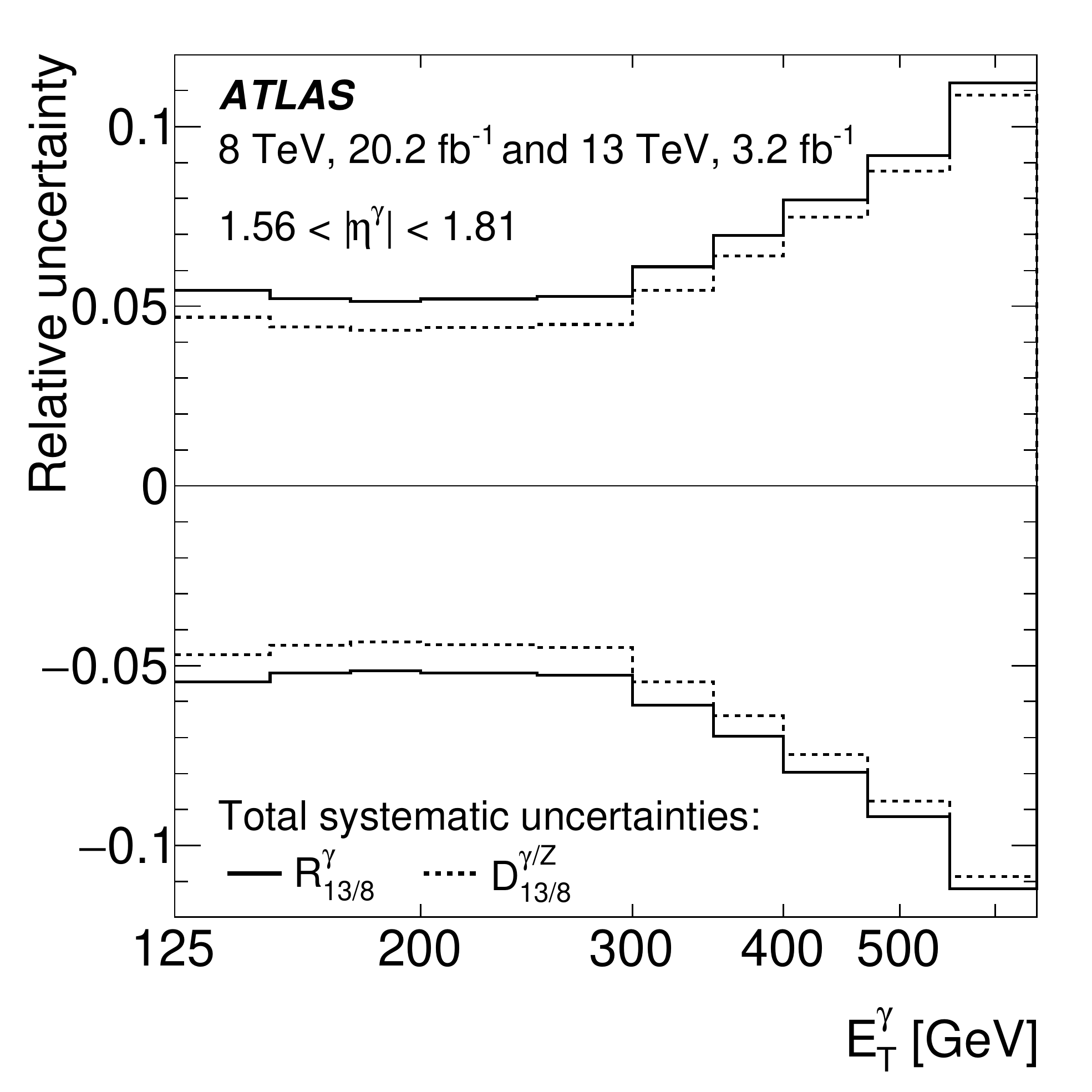}}
\put (8.0,0.0){\includegraphics[width=8cm]{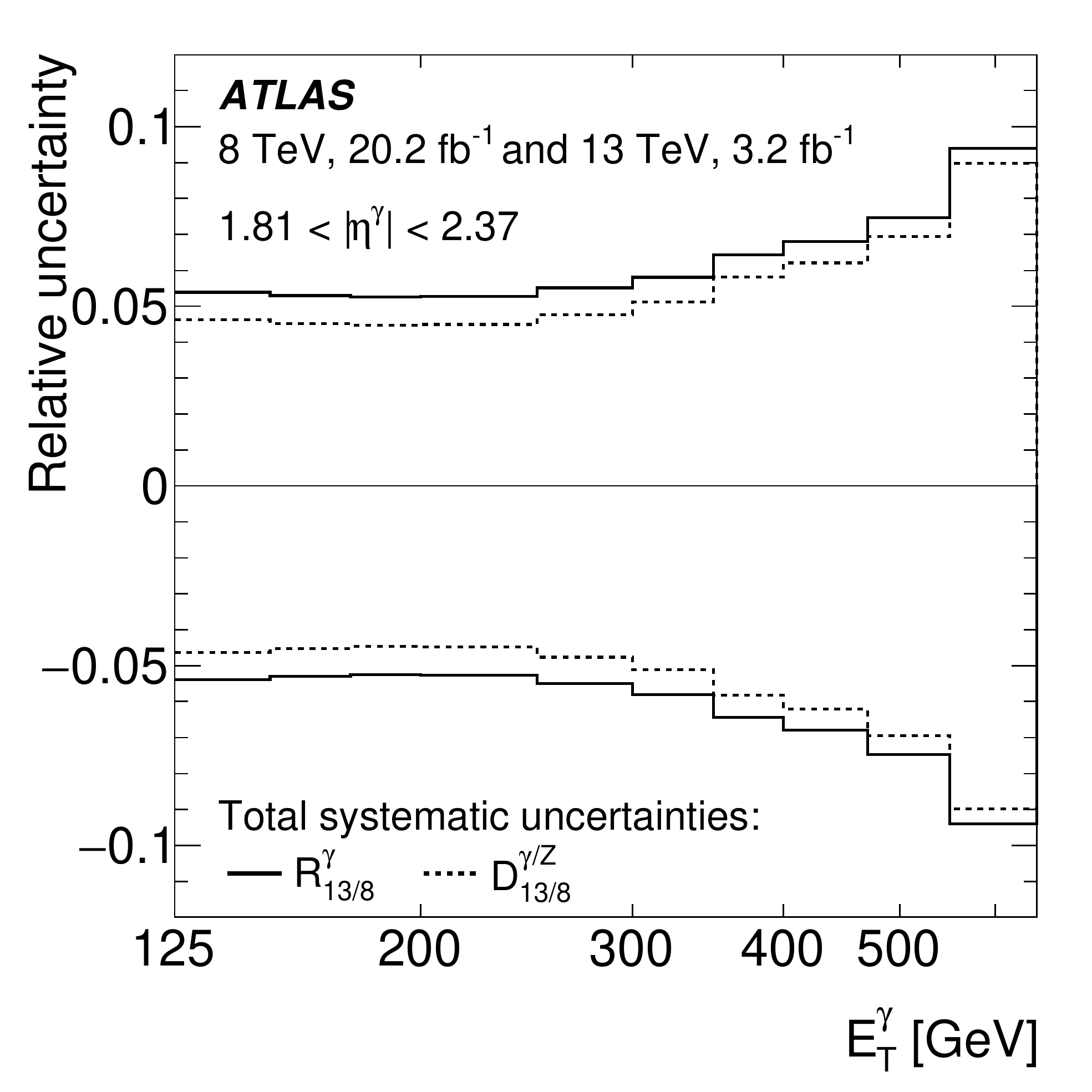}}
\end{picture}
\caption
{
  Relative total systematic uncertainty in $\R138$ (solid lines) and
  in $\D138$ (dashed lines) as functions of $\etg$ for different
  $\etag$ regions.
}
\label{fig04}
\end{figure}

The relative total systematic uncertainty in the measured $\D138$ as a
function of $\etg$ is shown for each $\etag$ region in
Figure~\ref{fig04}. For comparison, the relative total systematic
uncertainty in $\R138$ is also shown. Since the total systematic
uncertainty in $\Z138$ is at least a factor of three smaller
than the total systematic uncertainty in $\R138$, the effect of adding
in quadrature such a contribution has a small impact. On the other
hand, the luminosity uncertainty, which amounts to $2.8\%$ for $\R138$, 
cancels out in $\D138$ and this has a significant impact except at
high $\etg$, where the statistical uncertainty dominates.

\section{Results}
\label{results}

The measurements of the ratios of cross sections are presented and the
main features exhibited by the data are described. The theoretical
predictions are compared with the experimental results for both  $\R138$
and $\D138$.

\subsection{Results for $\R138$}
\label{resgratio}
The measured $\R138$ as a function of $\etg$ in different regions of
$|\etag|$ is shown in Figures~\ref{fig06} and \ref{fig07} and
Table~\ref{tableresratio}. The measured $\R138$ increases with $\etg$
from approximately $2$ at $\etg=125$~\GeV\ to
approximately $8$--$29$ at the high end of the spectrum. In the
forward regions the increase of $\R138$ with $\etg$ is larger than in
the central regions. At a fixed value of $\etg$, the measured ratio
increases as $|\etag|$ increases.

\begin{table}[h]
\caption{
 The measured $\R138$ as a function of $\etg$ together with the
 statistical uncertainty and total systematic uncertainty in different
 regions of $|\etag|$.
}
\setlength{\unitlength}{1.0cm}
\begin{picture} (18.0,8.0)
\put (0.0,3.5){\scalebox{1.0}{
\begin{tabular}{|c|l @{ $\pm$ } l @{ $\pm$ } l|l @{ $\pm$ } l @{ $\pm$ } l|l @{ $\pm$ } l @{ $\pm$ } l|l @{ $\pm$ } l @{ $\pm$ } l|} \hline
$\etg$ [\GeV]  & \multicolumn{12}{c|}{ $\R138 \pm {\mathrm{statistical\, uncertainty}} \pm {\mathrm{systematic\, uncertainty}}$ } \\ \hline
           &  \multicolumn{3}{c|}{$|\etag|<0.6$} &  \multicolumn{3}{c|}{$0.6 <|\etag|<1.37$} &   \multicolumn{3}{c|}{$1.56 <|\etag|<1.81$} &  \multicolumn{3}{c|}{$1.81 <|\etag|<2.37$} \\ \hline
$125$--$150$      &  $2.08   $ & $ 0.01     $ & $ 0.09$  &  $2.11 $ & $ 0.01   $ & $ 0.11$  &  $2.16 $ & $ 0.01 $ & $ 0.12$  &  $2.25 $ & $ 0.01   $ & $ 0.12$   \\  \hline
$150$--$175$      &  $2.12   $ & $ 0.01     $ & $ 0.08$  &  $2.15 $ & $ 0.01   $ & $ 0.10$  &  $2.22 $ & $ 0.02 $ & $ 0.12$  &  $2.46 $ & $ 0.05   $ & $ 0.13$   \\  \hline
$175$--$200$      &  $2.23   $ & $ 0.02     $ & $ 0.09$  &  $2.21 $ & $ 0.02   $ & $ 0.10$  &  $2.35 $ & $ 0.03 $ & $ 0.12$  &  $2.66 $ & $ 0.03   $ & $ 0.14$   \\  \hline
$200$--$250$      &  $2.28   $ & $ 0.02     $ & $ 0.09$  &  $2.28 $ & $ 0.02   $ & $ 0.10$  &  $2.63 $ & $ 0.03 $ & $ 0.14$  &  $3.10 $ & $ 0.03   $ & $ 0.16$   \\  \hline
$250$--$300$      &  $2.42   $ & $ 0.03     $ & $ 0.09$  &  $2.43 $ & $ 0.03   $ & $ 0.10$  &  $3.06 $ & $ 0.06 $ & $ 0.16$  &  $3.89 $ & $ 0.06   $ & $ 0.21$   \\  \hline
$300$--$350$      &  $2.53   $ & $ 0.04     $ & $ 0.10$  &  $2.72 $ & $ 0.04   $ & $ 0.12$  &  $3.67 $ & $ 0.12 $ & $ 0.22$  &  $5.18 $ & $ 0.15   $ & $ 0.30$   \\  \hline
$350$--$400$      &  $2.64   $ & $ 0.07     $ & $ 0.11$  &  $2.78 $ & $ 0.07   $ & $ 0.13$  &  $3.95 $ & $ 0.20 $ & $ 0.27$  &  $6.66 $ & $ 0.31   $ & $ 0.43$   \\  \hline
$400$--$470$      &  $2.83   $ & $ 0.09     $ & $ 0.11$  &  $3.11 $ & $ 0.09   $ & $ 0.15$  &  $5.73 $ & $ 0.35 $ & $ 0.46$  &  $8.43 $ & $ 0.55   $ & $ 0.57$   \\  \hline
$470$--$550$      &  $3.11   $ & $ 0.14     $ & $ 0.13$  &  $3.46 $ & $ 0.15   $ & $ 0.18$  &  $8.68 $ & $ 0.87 $ & $ 0.80$  &  $16.1 $ & $ 2.0     $ & $ 1.2$     \\  \hline
$550$--$650$      &  $3.28   $ & $ 0.21     $ & $ 0.14$  &  $4.35 $ & $ 0.27   $ & $ 0.24$  &  $12.5 $ & $ 2.3   $ & $ 1.4$    &  $29.3 $ & $ 7.5     $ & $ 2.8$     \\  \hline
$650$--$750$      &  $4.00   $ & $ 0.42     $ & $ 0.18$  &  $5.03 $ & $ 0.52   $ & $ 0.29$  & \multicolumn{6}{c|}{} \\ \hline
$750$--$900$      &  $5.20   $ & $ 0.75     $ & $ 0.25$  &  $8.4   $ & $ 1.3     $ & $ 0.5$    &  \multicolumn{6}{c|}{} \\ \hline
$900$--$1100$    &  $9.9     $ & $ 2.3       $ & $ 0.5$    &  $7.9   $ & $ 2.4     $ & $ 0.5$    &  \multicolumn{6}{c|}{} \\ \hline
$1100$--$1500$  &  $13.9   $ & $ 9.8       $ & $ 0.8$    &  \multicolumn{9}{c|}{}  \\ \hline
\end{tabular}
}}
\end{picture}
\label{tableresratio}
\end{table}

The NLO QCD predictions based on the MMHT2014 PDFs are compared with the
measured $\R138$ in Figures~\ref{fig06} and \ref{fig07}. 
Even though there is a tendency for the predictions to underestimate the data,
the measurements and the theory are consistent within the uncertainties;
in particular, the increase as $\etg$ increases and the dependence on $\etag$
are reproduced by the predictions. To study in more
detail the description of the measured $\R138$ by the NLO QCD
predictions, the ratio of the predictions to the data is shown in
Figures~\ref{fig06} and \ref{fig07}. In these figures, the predictions
based on different PDFs, namely MMHT2014, CT14, NNPDF3.0, HERAPDF2.0
and ABMP16 are included to ascertain the sensitivity of $\R138$ to the
proton PDFs. The predictions generally agree with the measured $\R138$
within the experimental and theoretical uncertainties for all PDFs
considered within the measured range. 

The comparison of the NLO QCD predictions for $\dsetg$ and the
measured differential  cross sections in the ATLAS analyses at $8$ and
$13$~\TeV\ is limited by the theoretical uncertainties, which are
larger than those of experimental nature and dominated by the
uncertainties due to the terms beyond NLO. The theoretical
uncertainties in $\dsetg$ are $10$--$15\%$; in contrast, the
theoretical uncertainties for $\R138$ are below $2\%$ for most of the
phase space considered and smaller than the experimental
uncertainties. The experimental uncertainties in $\R138$ also benefit
from a significant reduction since the systematic uncertainties partially
cancel out, in particular those related to the photon energy
scale, which is dominant in the measurement of $\dsetg$. The total
systematic uncertainty in $\R138$ is below $5\%$ for most of the phase
space considered. Thus, the significant reduction of the experimental
and theoretical uncertainties in $\R138$ allows a more stringent test
of NLO QCD. The overall level of agreement between data and the NLO QCD
predictions based on several parameterisations of the proton PDFs
within these reduced uncertainties validates the description of the
evolution of isolated-photon production in $pp$ collisions with the
centre-of-mass energy.

\begin{figure}[h]
\setlength{\unitlength}{1.0cm}
\begin{picture} (18.0,15.0)
\put (0.0,0.0){\includegraphics[width=8cm]{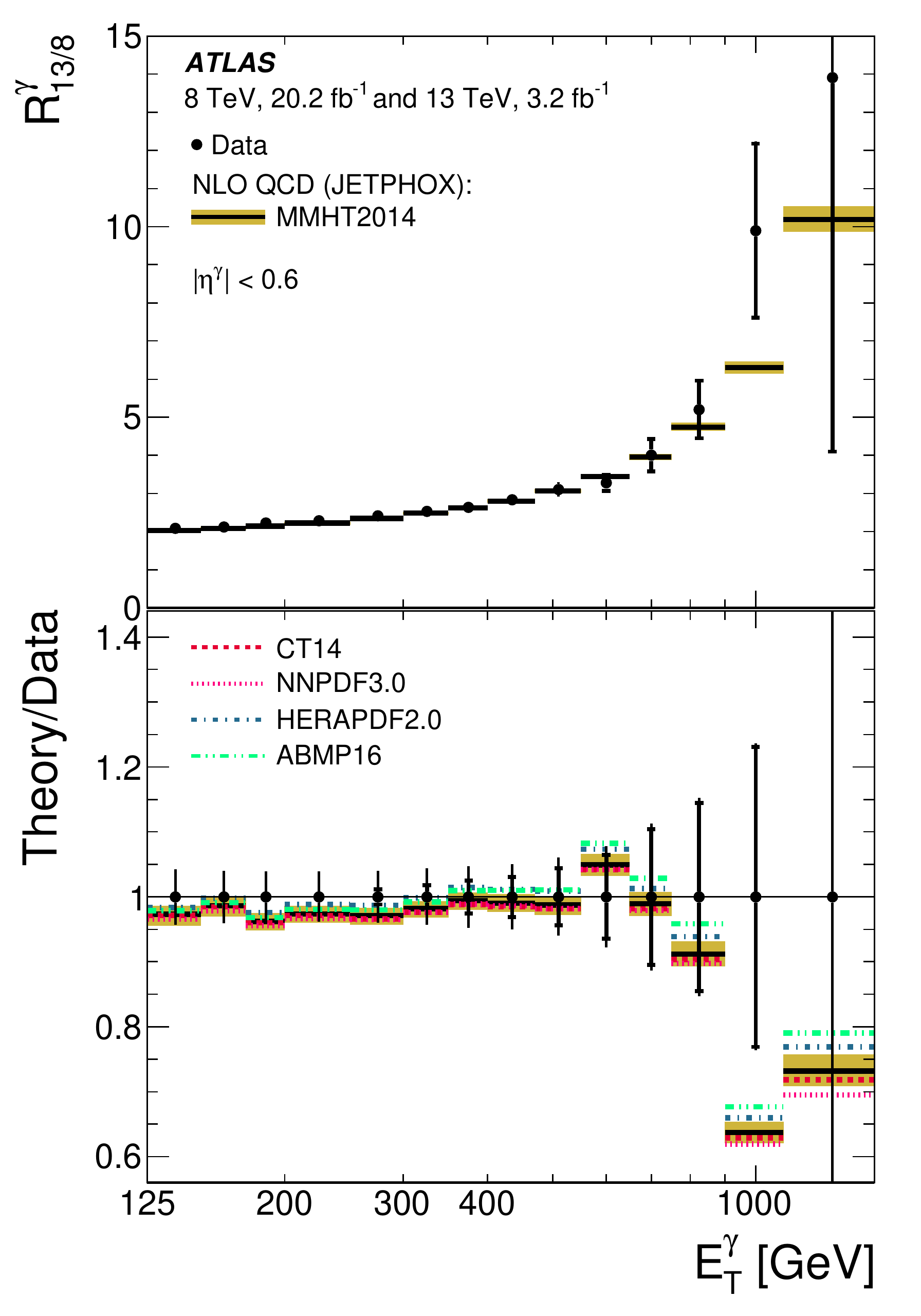}}
\put (8.0,0.0){\includegraphics[width=8cm]{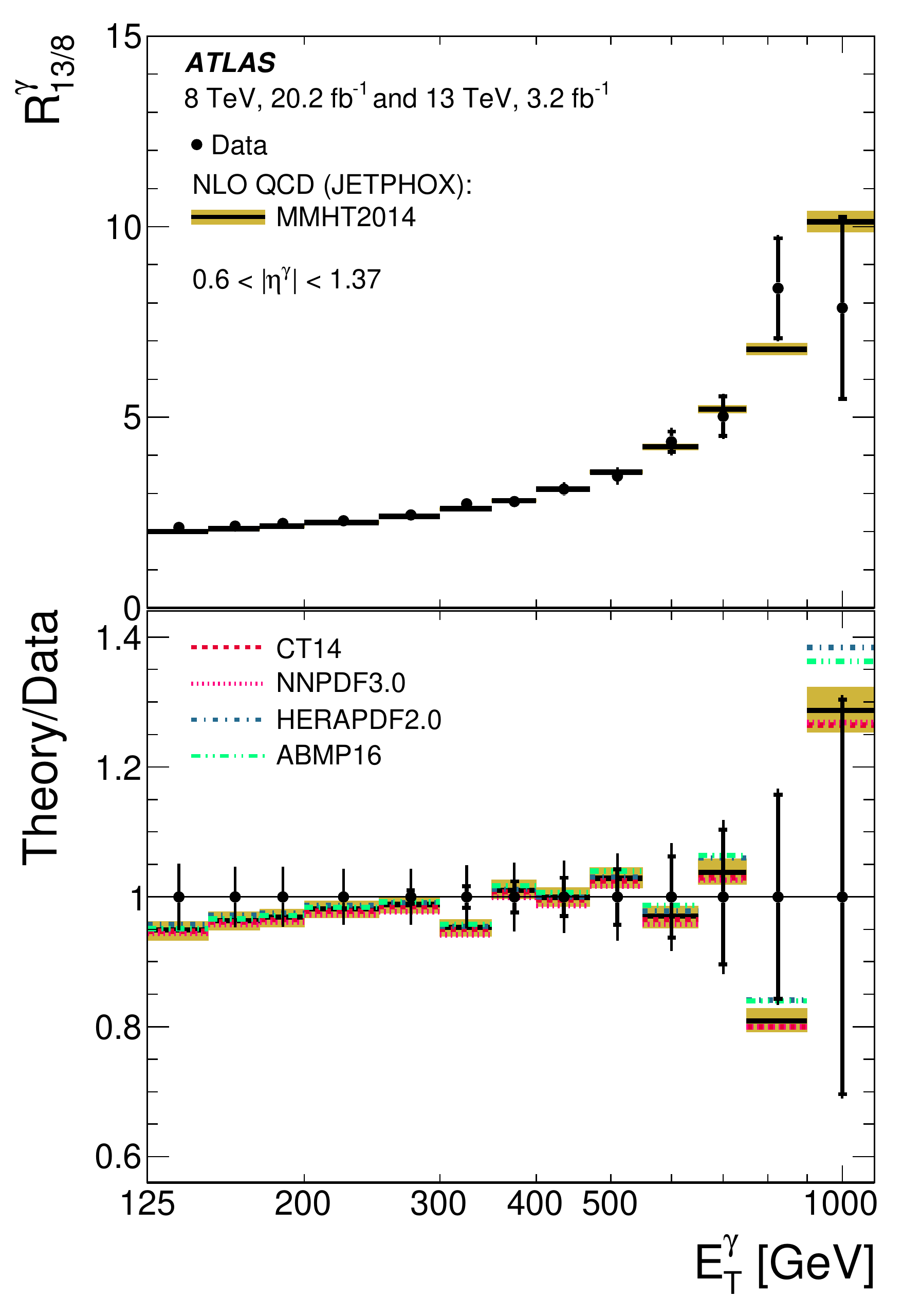}}
\end{picture}
\caption
{
  The measured $\R138$ (dots) as a function of $\etg$ in different
  regions of $|\etag|$. The NLO QCD predictions based on the MMHT2014
  PDFs (black lines) are also shown. The inner (outer) error bars
  represent the statistical (total) uncertainties. The shaded band
  represents the theoretical uncertainty in the predictions. For most
  of the points, the error bars are smaller than the marker size and,
  thus, not visible. The lower part of the figures shows the ratio of
  the NLO QCD predictions based on the MMHT2014 PDFs to the measured
  $\R138$ (black lines). The ratios of the NLO QCD predictions based
  on different PDF sets to the measured $\R138$ are also included.
}
\label{fig06}
\end{figure}

\begin{figure}[h]
\setlength{\unitlength}{1.0cm}
\begin{picture} (18.0,16.0)
\put (0.0,0.0){\includegraphics[width=8cm]{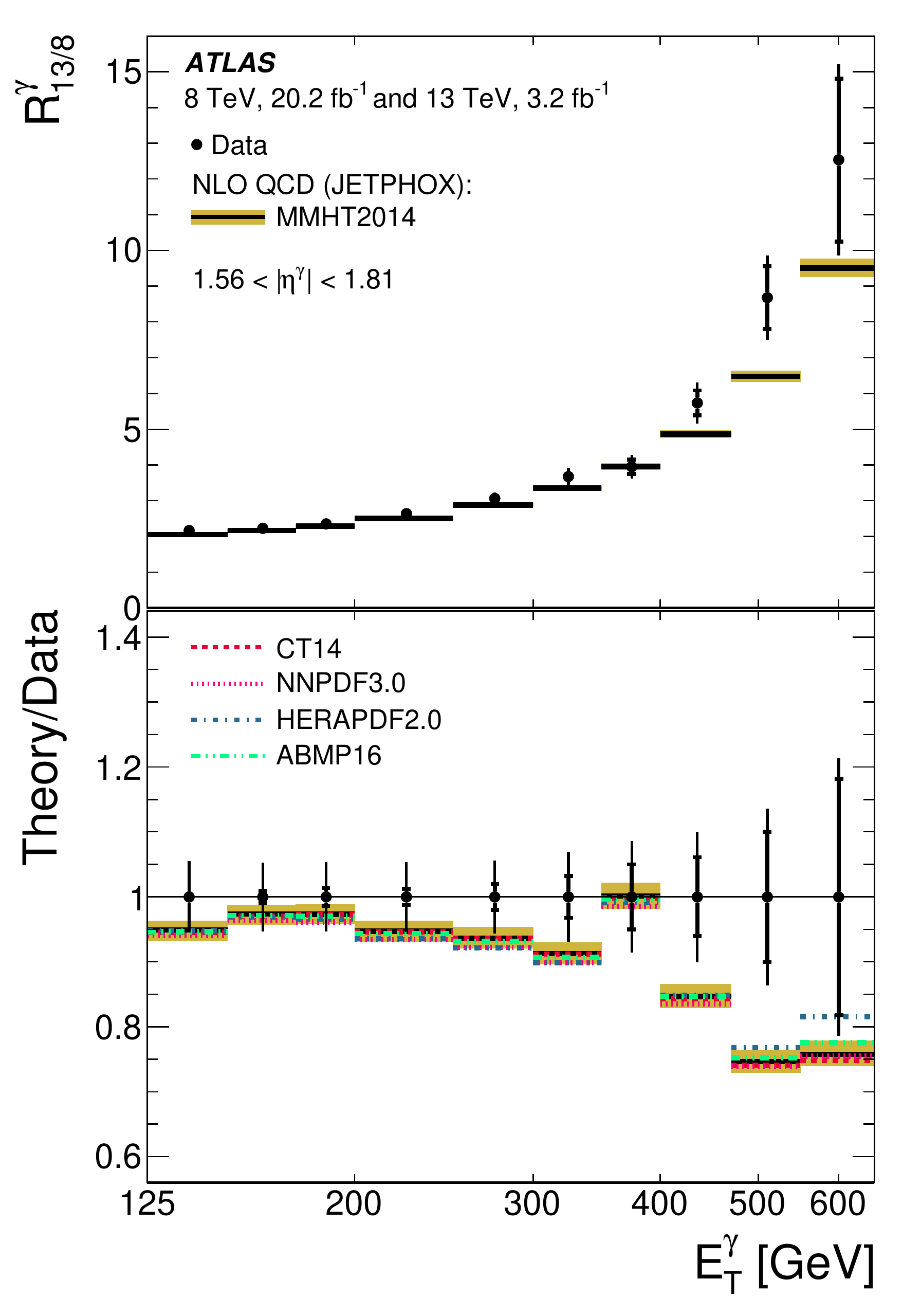}}
\put (8.0,0.0){\includegraphics[width=8cm]{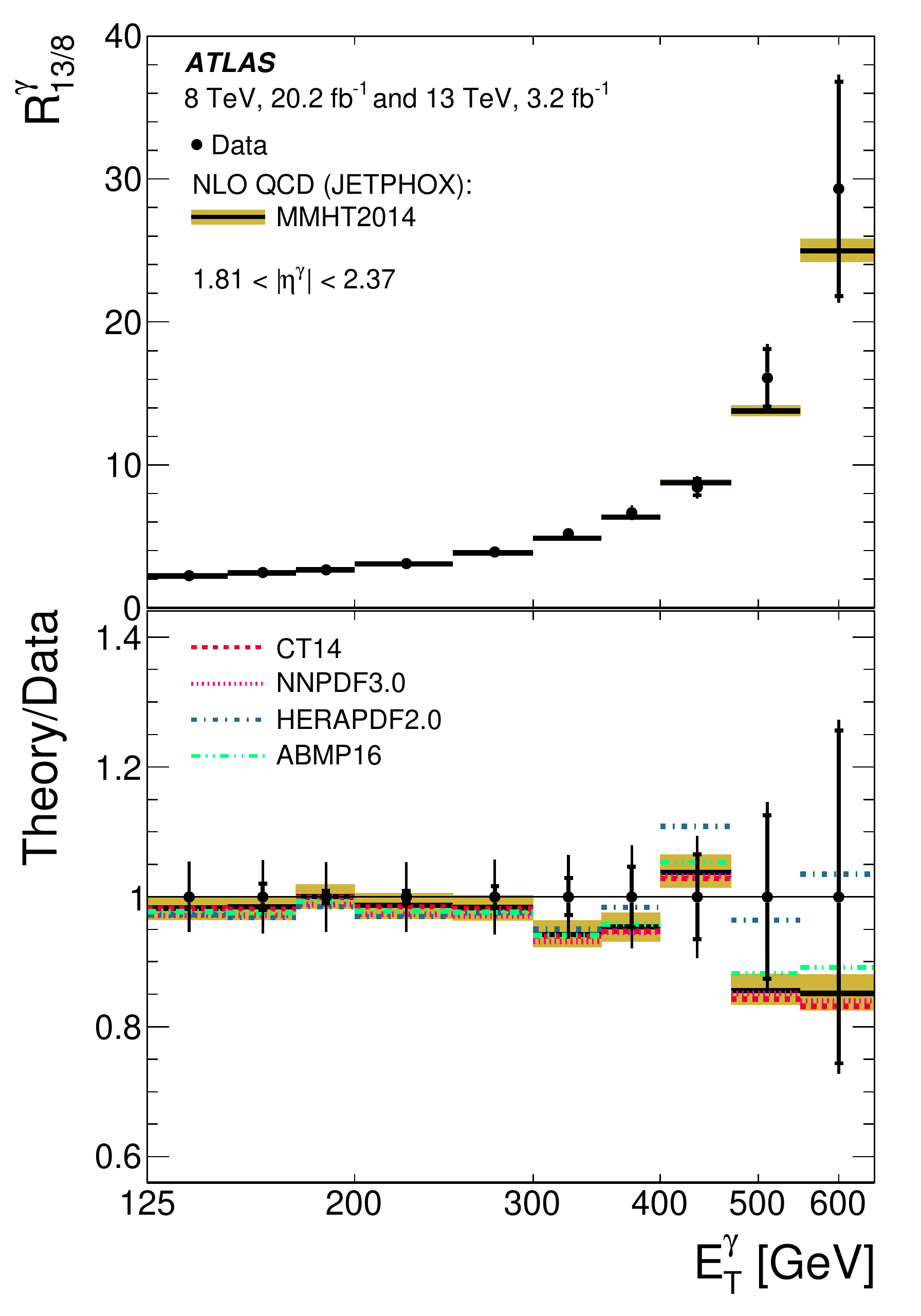}}
\end{picture}
\caption
{
  The measured $\R138$ (dots) as a function of $\etg$ in different
  regions of $|\etag|$. The NLO QCD predictions based on the MMHT2014
  PDFs (black lines) are also shown. The inner (outer) error bars
  represent the statistical (total) uncertainties. The shaded band
  represents the theoretical uncertainty in the predictions. For most
  of the points, the error bars are smaller than the marker size and,
  thus, not visible. The lower part of the figures shows the ratio of
  the NLO QCD predictions based on the MMHT2014 PDFs to the measured
  $\R138$ (black lines). The ratios of the NLO QCD predictions based
  on different PDF sets to the measured $\R138$ are also included.
}
\label{fig07}
\end{figure}

\FloatBarrier
\subsection{Results for $\D138$}
\label{reszratio}
The measurements of $\D138$ as a function of $\etg$ in different regions of
$|\etag|$ are shown in Figures~\ref{fig08} and \ref{fig09} and
Table~\ref{tablereszratio}. 
The measured $\D138$ increases with
$\etg$ from approximately $1.4$ at $\etg=125$~\GeV\ to
approximately $5$--$19$ at the high end of the spectrum. At a fixed
value of $\etg$, the measured ratio increases as $|\etag|$ increases.

The theoretical predictions based on the MMHT2014nnlo PDFs are
compared with the measured $\D138$ in Figures~\ref{fig08} and
\ref{fig09}. The predictions are in agreement with the measured
$\D138$; in particular, the increase as $\etg$ increases and
the dependence on $\etag$ are reproduced by the predictions. As an
example, the measured value of $\D138$ at the lowest-$\etg$ point for
$|\etag|<0.6$ is $1.35 \pm 0.04$ while the prediction using MMHT2014
is $1.31 \pm 0.02$. The tendency of the predictions to underestimate
the data observed in $\R138$ is also present in $\D138$; nevertheless,
they are still consistent with each other within the uncertainties. To
study in more detail the description of the measured $\D138$ by the
theoretical predictions, the ratio of the predictions to the data is
shown in Figures~\ref{fig08} and \ref{fig09}. In these figures, the
predictions based on different PDFs, namely MMHT2014nnlo, CT14nnlo,
NNPDF3.0nnlo and HERAPDF2.0nnlo are included to estimate the
sensitivity of $\D138$ to the proton PDFs. The predictions generally
agree with the measured $\D138$ within the experimental and
theoretical uncertainties for all PDFs considered within the measured
range.

\begin{table}[h]
\caption{
 The measured $\D138$ as a function of $\etg$ together with the
 statistical and total systematic uncertainty in different regions of
 $|\etag|$.
}
\setlength{\unitlength}{1.0cm}
\begin{picture} (18.0,8.0)
\put (0.0,3.5){\scalebox{1.0}{
\begin{tabular}{|c|l @{ $\pm$ } l @{ $\pm$ } l|l @{ $\pm$ } l @{ $\pm$ } l|l @{ $\pm$ } l @{ $\pm$ } l|l @{ $\pm$ } l @{ $\pm$ } l|} \hline
$\etg$ [\GeV] & \multicolumn{12}{c|}{ $\D138 \pm {\mathrm{statistical\, uncertainty}} \pm {\mathrm{systematic\, uncertainty}}$ } \\ \hline
           &  \multicolumn{3}{c|}{$|\etag|<0.6$} &  \multicolumn{3}{c|}{$0.6 <|\etag|<1.37$} &   \multicolumn{3}{c|}{$1.56 <|\etag|<1.81$} &  \multicolumn{3}{c|}{$1.81 <|\etag|<2.37$} \\ \hline
$125$--$150$      &  $1.35 $ & $ 0.01    $ & $ 0.04$  & $1.37   $ & $ 0.00   $ & $ 0.06$ &  $1.40   $ & $ 0.01     $ & $ 0.07$  & $1.46   $ & $ 0.01     $ & $ 0.07$  \\  \hline
$150$--$175$      &  $1.38 $ & $ 0.01    $ & $ 0.04$  & $1.40   $ & $ 0.01   $ & $ 0.05$ &  $1.44   $ & $ 0.01     $ & $ 0.06$  & $1.60   $ & $ 0.03     $ & $ 0.07$  \\  \hline
$175$--$200$      &  $1.45 $ & $ 0.01    $ & $ 0.04$  & $1.44   $ & $ 0.01   $ & $ 0.05$ &  $1.53   $ & $ 0.02     $ & $ 0.07$  & $1.73   $ & $ 0.02     $ & $ 0.08$  \\  \hline
$200$--$250$      &  $1.49 $ & $ 0.01    $ & $ 0.04$  & $1.49   $ & $ 0.01   $ & $ 0.05$ &  $1.71   $ & $ 0.02     $ & $ 0.08$  & $2.02   $ & $ 0.02     $ & $ 0.09$  \\  \hline
$250$--$300$      &  $1.57 $ & $ 0.02    $ & $ 0.04$  & $1.58   $ & $ 0.02   $ & $ 0.05$ &  $1.99   $ & $ 0.04     $ & $ 0.09$  & $2.53   $ & $ 0.04     $ & $ 0.12$  \\  \hline
$300$--$350$      &  $1.65 $ & $ 0.03    $ & $ 0.05$  & $1.77   $ & $ 0.03   $ & $ 0.06$ &  $2.39   $ & $ 0.08     $ & $ 0.13$  & $3.37   $ & $ 0.10     $ & $ 0.17$  \\  \hline
$350$--$400$      &  $1.72 $ & $ 0.04    $ & $ 0.05$  & $1.81   $ & $ 0.04   $ & $ 0.07$ &  $2.57   $ & $ 0.13     $ & $ 0.16$  & $4.33   $ & $ 0.20     $ & $ 0.25$  \\  \hline
$400$--$470$      &  $1.84 $ & $ 0.06    $ & $ 0.05$  & $2.02   $ & $ 0.06   $ & $ 0.08$ &  $3.73   $ & $ 0.23     $ & $ 0.28$  & $5.48   $ & $ 0.36     $ & $ 0.34$  \\  \hline
$470$--$550$      &  $2.02 $ & $ 0.09    $ & $ 0.06$  & $2.25   $ & $ 0.10   $ & $ 0.10$ &  $5.65   $ & $ 0.57     $ & $ 0.50$  & $10.5   $ & $ 1.3       $ & $ 0.7$    \\  \hline
$550$--$650$      &  $2.13 $ & $ 0.14    $ & $ 0.07$  & $2.83   $ & $ 0.18   $ & $ 0.14$ &  $8.2     $ & $ 1.5       $ & $ 0.9$    & $19.1   $ & $ 4.9       $ & $ 1.7$    \\  \hline
$650$--$750$      &  $2.60 $ & $ 0.27    $ & $ 0.09$  & $3.27   $ & $ 0.34   $ & $ 0.17$ &  \multicolumn{6}{c|}{} \\ \hline
$750$--$900$      &  $3.39 $ & $ 0.49    $ & $ 0.14$  & $5.46   $ & $ 0.86   $ & $ 0.27$ &  \multicolumn{6}{c|}{} \\ \hline
$900$--$1100$    &  $6.4   $ & $ 1.5      $ & $ 0.3$    & $5.1     $ & $ 1.6     $ & $ 0.3$ &  \multicolumn{6}{c|}{} \\ \hline
$1100$--$1500$  &  $9.1   $ & $ 6.4      $ & $ 0.5$    &  \multicolumn{9}{c|}{}  \\ \hline
\end{tabular}
}}
\end{picture}
\label{tablereszratio}
\end{table}

\begin{figure}[h]
\setlength{\unitlength}{1.0cm}
\begin{picture} (18.0,16.0)
\put (0.0,0.0){\includegraphics[width=8cm]{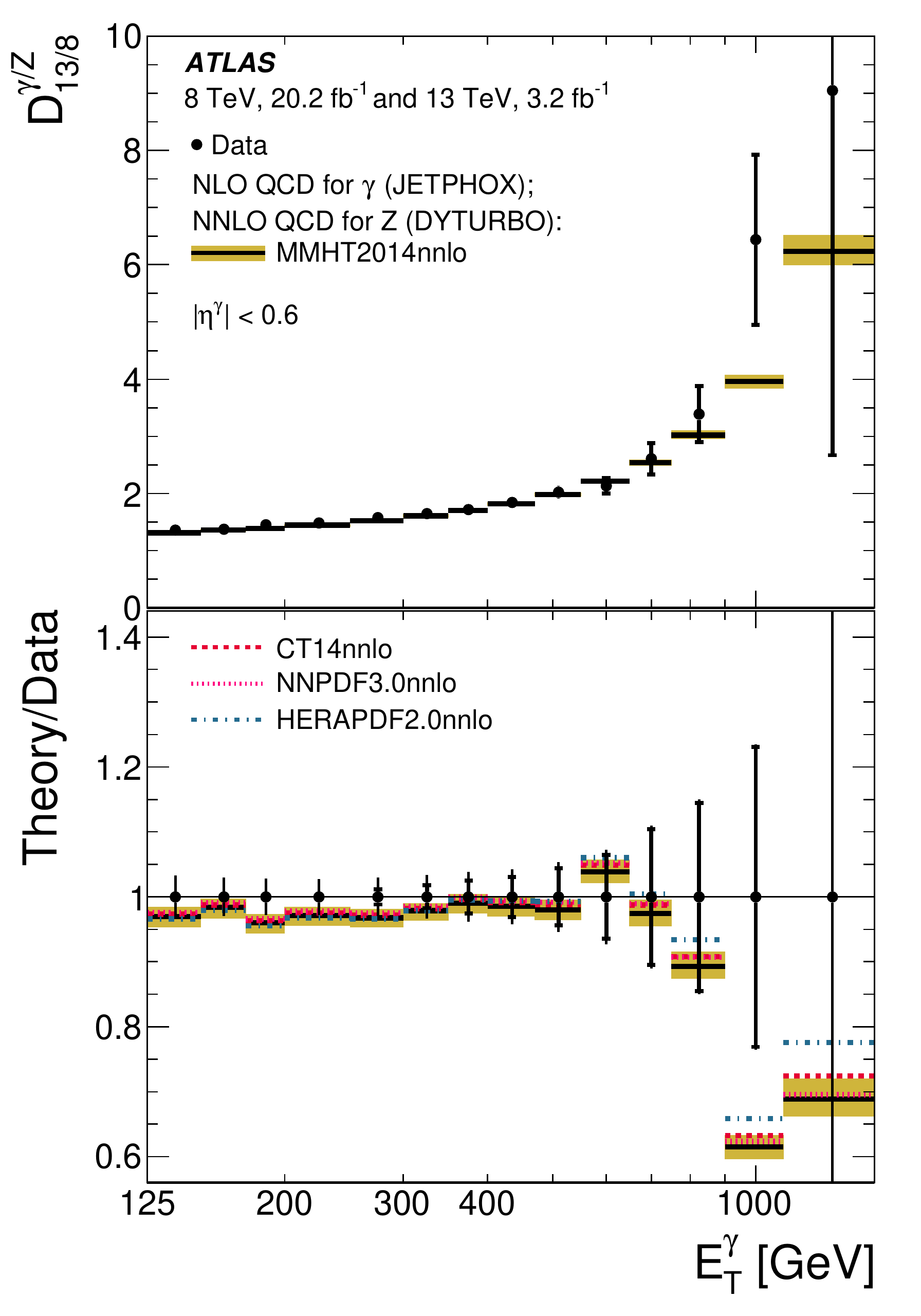}}
\put (8.0,0.0){\includegraphics[width=8cm]{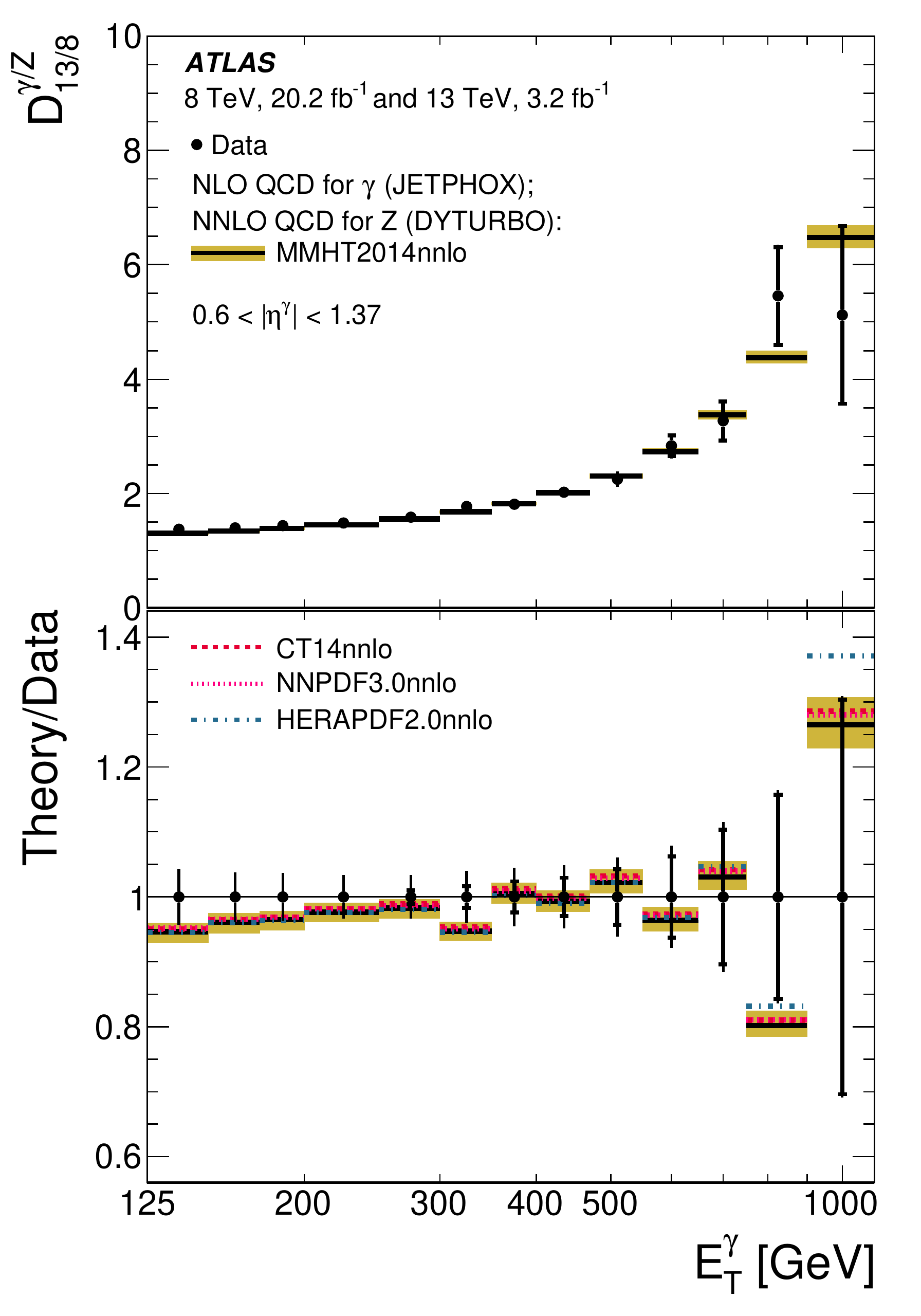}}
\end{picture}
\caption
{
  The measured $\D138$ (dots) as a function of $\etg$ in different
  regions of $|\etag|$. The pQCD predictions based on the MMHT2014nnlo
  PDFs (black lines) are also shown. The inner (outer) error bars
  represent the statistical (total) uncertainties. The shaded band
  represents the theoretical uncertainty in the predictions. For most
  of the points, the error bars are smaller than the marker size and,
  thus, not visible. The lower part of the figures shows the ratio of
  the pQCD predictions based on the MMHT2014nnlo PDFs to the measured
  $\D138$ (black lines). The ratios of the pQCD predictions based on
  different PDF sets to the measured $\D138$ are also included.
}
\label{fig08}
\end{figure}

\begin{figure}[h]
\setlength{\unitlength}{1.0cm}
\begin{picture} (18.0,16.0)
\put (0.0,0.0){\includegraphics[width=8cm]{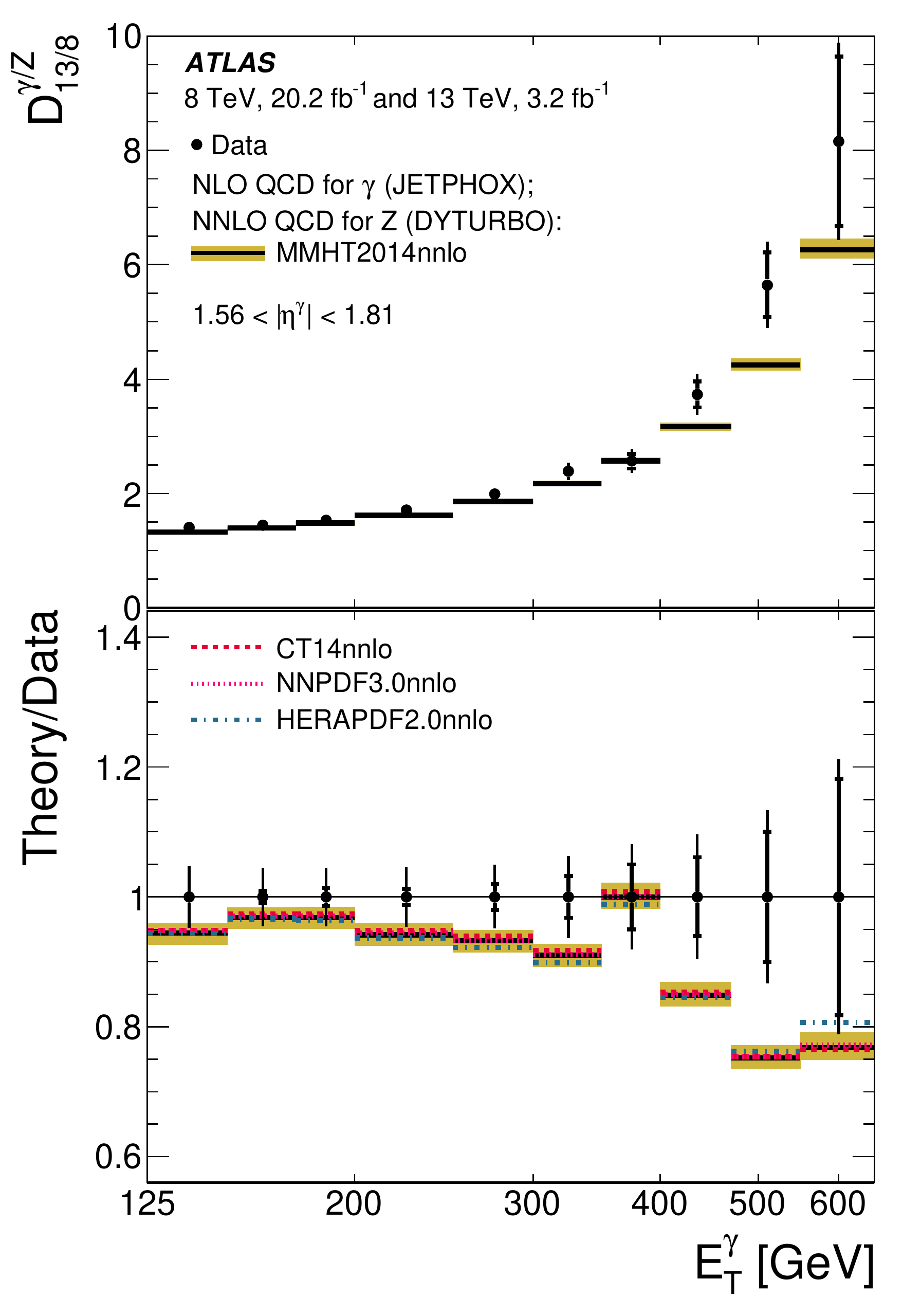}}
\put (8.0,0.0){\includegraphics[width=8cm]{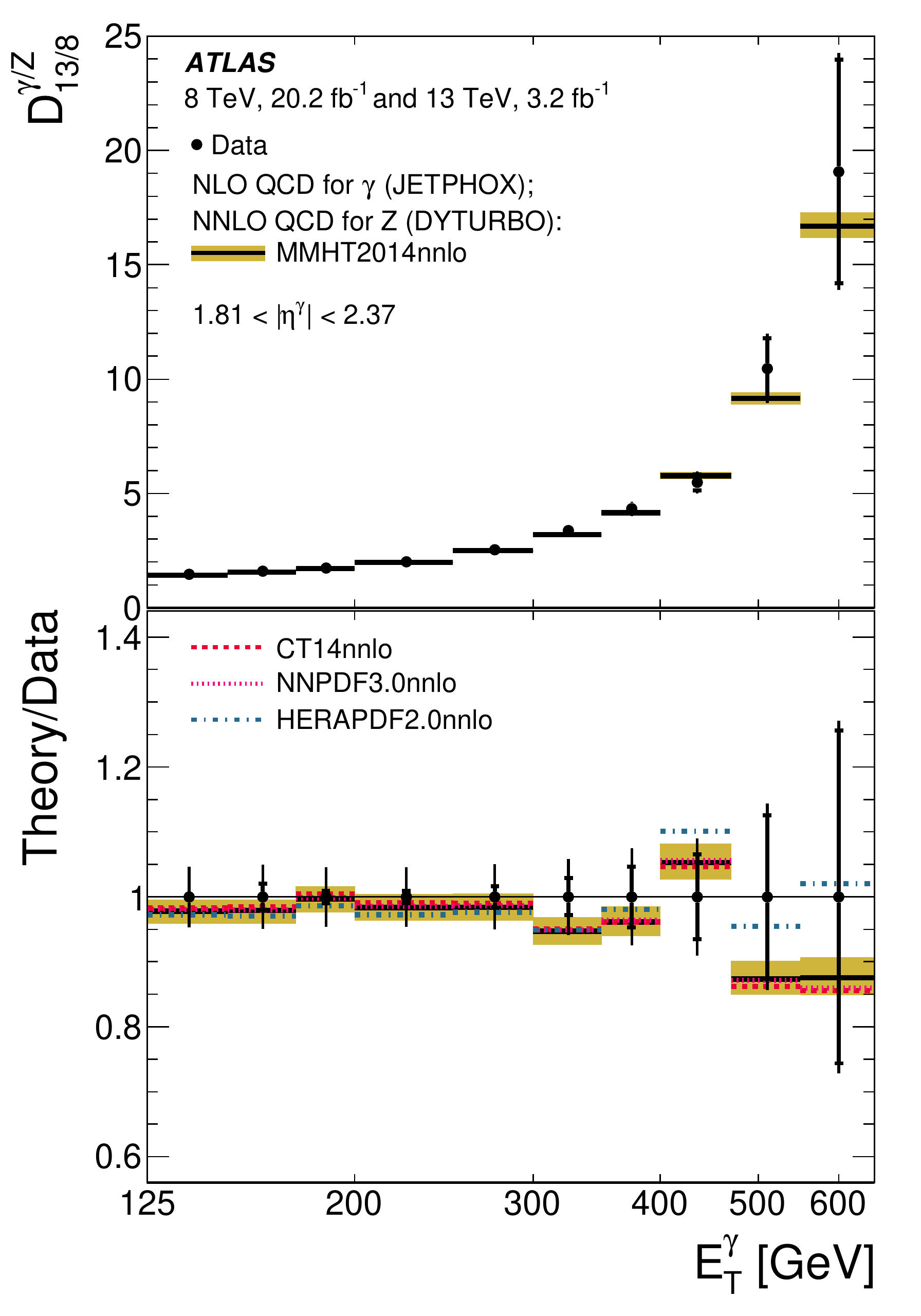}}
\end{picture}
\caption
{
  The measured $\D138$ (dots) as a function of $\etg$ in different
  regions of $|\etag|$. The pQCD predictions based on the MMHT2014nnlo
  PDFs (black lines) are also shown. The inner (outer) error bars
  represent the statistical (total) uncertainties. The shaded band
  represents the theoretical uncertainty in the predictions. For most
  of the points, the error bars are smaller than the marker size and,
  thus, not visible. The lower part of the figures shows the ratio of
  the pQCD predictions based on the MMHT2014nnlo PDFs to the measured
  $\D138$ (black lines). The ratios of the pQCD predictions based on
  different PDF sets to the measured $\D138$ are also included.
}
\label{fig09}
\end{figure}

\FloatBarrier
\section{Summary and conclusions}
\label{summary}
The ratio of cross sections for inclusive isolated-photon production in $pp$ 
collisions at $\sqrt s=13$ and $8$~\TeV\ ($\R138$) is measured using the 
ATLAS detector at the LHC. The integrated luminosities of the 13~\TeV\ and 
8~\TeV\ datasets are 3.2~\fb1\ and 20.2~\fb1, respectively.
The ratio of differential cross sections as a function of $\etg$ is
measured in different regions of $|\etag|$ for photons with $125< \etg
<1500$~\GeV\ and $|\etag|<2.37$, excluding the region
$1.37<|\etag|<1.56$. 
In the estimation of the experimental systematic uncertainties for
$\R138$, the correlations between the measurements at the two
centre-of-mass energies are taken into account. The systematic
uncertainty arising from the photon energy scale, which is dominant
for the individual cross sections, is reduced significantly in $\R138$
and no longer the dominant uncertainty. The total systematic
uncertainty for $\R138$ is below $5\%$ in most of the phase space of
the measurement. The measurements can be useful for tuning models of
prompt-photon production in $pp$ collisions.

The predictions from NLO QCD calculations are compared with the measured
$\R138$. The theoretical uncertainties affecting these predictions are
also evaluated taking into account the correlations between the two
centre-of-mass energies, resulting in   a significant reduction in the
uncertainty of the predicted $\R138$. The theoretical uncertainties in
$\R138$ are below $2\%$ for most of the phase space of the
measurement, in contrast with those in the individual cross-section
predictions, which have      
approximately $10$--$15\%$ uncertainties. Thus, the
comparison of the predictions with the measured $\R138$ represents a
stringent test of the pQCD calculations. Within these reduced
experimental and theoretical uncertainties, the NLO QCD predictions
based on several parameterisations of the proton PDFs agree with the
data. Even though there is a tendency of the predictions to
underestimate the data, the measurements and the theory are consistent
within the uncertainties. The level of agreement achieved validates
the description of the evolution of isolated-photon production in $pp$
collisions from $\sqrt s=8$ to $13$~\TeV.

A double ratio of cross sections is also measured: the ratio of $\R138$ to
the ratio of the fiducial cross sections for $Z$ boson production at
$13$ and $8$~\TeV\ ($\Z138$). In $\D138\equiv\R138/\Z138$, the
uncertainty due to the luminosity cancels out at the expense of a
small increase in the systematic uncertainty from all other sources,
leading to a more precise measurement of the evolution of the
inclusive-photon cross section with the centre-of-mass energy
normalised to the evolution of the $Z$ boson cross section. The
theoretical prediction, based on NNLO (NLO) QCD calculations for
$Z$ boson (inclusive-photon) production, describes the measurements
within the theoretical uncertainties and the reduced experimental
uncertainties.

\section*{Acknowledgements}


We thank CERN for the very successful operation of the LHC, as well as the
support staff from our institutions without whom ATLAS could not be
operated efficiently.

We acknowledge the support of ANPCyT, Argentina; YerPhI, Armenia; ARC, Australia; BMWFW and FWF, Austria; ANAS, Azerbaijan; SSTC, Belarus; CNPq and FAPESP, Brazil; NSERC, NRC and CFI, Canada; CERN; CONICYT, Chile; CAS, MOST and NSFC, China; COLCIENCIAS, Colombia; MSMT CR, MPO CR and VSC CR, Czech Republic; DNRF and DNSRC, Denmark; IN2P3-CNRS, CEA-DRF/IRFU, France; SRNSFG, Georgia; BMBF, HGF, and MPG, Germany; GSRT, Greece; RGC, Hong Kong SAR, China; ISF and Benoziyo Center, Israel; INFN, Italy; MEXT and JSPS, Japan; CNRST, Morocco; NWO, Netherlands; RCN, Norway; MNiSW and NCN, Poland; FCT, Portugal; MNE/IFA, Romania; MES of Russia and NRC KI, Russian Federation; JINR; MESTD, Serbia; MSSR, Slovakia; ARRS and MIZ\v{S}, Slovenia; DST/NRF, South Africa; MINECO, Spain; SRC and Wallenberg Foundation, Sweden; SERI, SNSF and Cantons of Bern and Geneva, Switzerland; MOST, Taiwan; TAEK, Turkey; STFC, United Kingdom; DOE and NSF, United States of America. In addition, individual groups and members have received support from BCKDF, CANARIE, CRC and Compute Canada, Canada; COST, ERC, ERDF, Horizon 2020, and Marie Sk{\l}odowska-Curie Actions, European Union; Investissements d' Avenir Labex and Idex, ANR, France; DFG and AvH Foundation, Germany; Herakleitos, Thales and Aristeia programmes co-financed by EU-ESF and the Greek NSRF, Greece; BSF-NSF and GIF, Israel; CERCA Programme Generalitat de Catalunya, Spain; The Royal Society and Leverhulme Trust, United Kingdom. 

The crucial computing support from all WLCG partners is acknowledged gratefully, in particular from CERN, the ATLAS Tier-1 facilities at TRIUMF (Canada), NDGF (Denmark, Norway, Sweden), CC-IN2P3 (France), KIT/GridKA (Germany), INFN-CNAF (Italy), NL-T1 (Netherlands), PIC (Spain), ASGC (Taiwan), RAL (UK) and BNL (USA), the Tier-2 facilities worldwide and large non-WLCG resource providers. Major contributors of computing resources are listed in Ref.~\cite{ATL-GEN-PUB-2016-002}.

\printbibliography

\newpage
 
\begin{flushleft}
{\Large The ATLAS Collaboration}

\bigskip

M.~Aaboud$^\textrm{\scriptsize 35d}$,    
G.~Aad$^\textrm{\scriptsize 101}$,    
B.~Abbott$^\textrm{\scriptsize 128}$,    
D.C.~Abbott$^\textrm{\scriptsize 102}$,    
O.~Abdinov$^\textrm{\scriptsize 13,*}$,    
D.K.~Abhayasinghe$^\textrm{\scriptsize 93}$,    
S.H.~Abidi$^\textrm{\scriptsize 167}$,    
O.S.~AbouZeid$^\textrm{\scriptsize 40}$,    
N.L.~Abraham$^\textrm{\scriptsize 156}$,    
H.~Abramowicz$^\textrm{\scriptsize 161}$
% ...
\bigskip
\\

% ...
$^{13}$Institute of Physics, Azerbaijan Academy of Sciences, Baku; Azerbaijan.\\
% ...
$^{35}$$^{(a)}$Facult\'e des Sciences Ain Chock, R\'eseau Universitaire de Physique des Hautes Energies - Universit\'e Hassan II, Casablanca;$^{(b)}$Facult\'{e} des Sciences, Universit\'{e} Ibn-Tofail, K\'{e}nitra;$^{(c)}$Facult\'e des Sciences Semlalia, Universit\'e Cadi Ayyad, LPHEA-Marrakech;$^{(d)}$Facult\'e des Sciences, Universit\'e Mohamed Premier and LPTPM, Oujda;$^{(e)}$Facult\'e des sciences, Universit\'e Mohammed V, Rabat; Morocco.\\
% ...
$^{40}$Niels Bohr Institute, University of Copenhagen, Copenhagen; Denmark.\\
% ...
$^{93}$Department of Physics, Royal Holloway University of London, Egham; United Kingdom.\\
% ...
$^{101}$CPPM, Aix-Marseille Universit\'e, CNRS/IN2P3, Marseille; France.\\
$^{102}$Department of Physics, University of Massachusetts, Amherst MA; United States of America.\\
% ...
$^{128}$Homer L. Dodge Department of Physics and Astronomy, University of Oklahoma, Norman OK; United States of America.\\
% ...
$^{156}$Department of Physics and Astronomy, University of Sussex, Brighton; United Kingdom.\\
% ...
$^{161}$Raymond and Beverly Sackler School of Physics and Astronomy, Tel Aviv University, Tel Aviv; Israel.\\
% ...
$^{167}$Department of Physics, University of Toronto, Toronto ON; Canada.\\
% ...
$^{a}$ Also at Borough of Manhattan Community College, City University of New York, New York NY; United States of America.\\
$^{b}$ Also at Centre for High Performance Computing, CSIR Campus, Rosebank, Cape Town; South Africa.\\
$^{c}$ Also at CERN, Geneva; Switzerland.\\
$^{d}$ Also at CPPM, Aix-Marseille Universit\'e, CNRS/IN2P3, Marseille; France.\\
$^{e}$ Also at D\'epartement de Physique Nucl\'eaire et Corpusculaire, Universit\'e de Gen\`eve, Gen\`eve; Switzerland.\\
% ...
$^{*}$ Deceased

\end{flushleft}


\end{document}